\documentclass{egpubl}

\ConferenceSubmission % uncomment for Conference submission
 \electronicVersion % can be used both for the printed and electronic version

\ifpdf \usepackage[pdftex]{graphicx} \pdfcompresslevel=9
\else \usepackage[dvips]{graphicx} \fi

\PrintedOrElectronic

\usepackage{t1enc,dfadobe}

\usepackage{egweblnk}
\usepackage{cite}

\usepackage{dsfont}
\usepackage{amsmath}
\usepackage{multirow}
\usepackage{xcolor}

\title{360$^\circ$ Panorama Cloning on Sphere}

\author[Q. Zhao et al.]
       {Qiang Zhao$^{1,2}$, Liang Wan$^{2}$\thanks{lwan@tju.edu.cn}, Wei Feng$^3$, Jiawan Zhang$^2$, Tien-Tsin Wong$^4$
        \\
         $^1$Institute of Computing Technology, Chinese Academy of Sciences, China \\
         $^2$School of Computer Software, Tianjin University, China \\
         $^3$School of Computer Science and Technology, Tianjin University, China \\
         $^4$Dept. of Computer Science and Engineering, The Chinese University of Hong Kong
       }

\begin{document}

% \teaser{
%  \includegraphics[width=\linewidth]{eg_new}
%  \centering
%   \caption{New EG Logo}
% \label{fig:teaser}
% }

\maketitle

\begin{abstract}
In this paper, we address a novel problem of cloning a patch of the source spherical panoramic image to the target spherical panoramic image, which we call 360$^\circ$ panorama cloning. Considering the sphere geometry constraint embedded in spherical panoramic images, we develop a coordinate-based method that directly clones in the spherical domain. Our method neither differentiates the polar regions and equatorial regions, nor identifies the boundaries in the unrolled planar-formatted panorama. We discuss in depth two unique issues in panorama cloning, i.e. preserving the patch's orientation, and handling the large-patch cloning (covering over 180$^\circ$ field-of-view) which may suffer from discoloration artifacts. As experimental results demonstrate, our method is able to get visually pleasing cloning results and achieve real time cloning performance.


\begin{classification} % according to http://www.acm.org/class/1998/
\CCScat{Computer Graphics}{I.3.3}{Picture/Image Generation}{}
\end{classification}

\end{abstract}

%-------------------------------------------------------------------------

\section{Introduction}

Panorama, a wide-angle view representation of a physical scene, has a long history dated back to the 18th century. Rather than being artistic works in early days, panoramic images have became more prevalent in recent years due to the progress of image stitching techniques and spherical imaging system, for displaying landscapes, street side views, or astronomic sceneries. Nowadays, enormous panoramas are publicly available via multiple well-known navigation websites, including Google Street View, Microsoft Bing Map, 360 Cities, etc. The public panoramas usually capture the whole field of view in all directions around the photographer, and most of which also cover up and down views. Due to the embedded sphere geometry, for viewing convenience a 360$^\circ$ panorama is commonly represented in the unrolled latitude-longitude planar format. It shows all the views of the panorama in a way natural to viewers, often with the horizon aligned to the equatorial line in the sphere, yet subjective to increasing distortions near two poles.

Recently there are a number of panorama-based applications emerging, including online street-level virtual navigation \cite{StreetView,Cube2Video13}, city-scale change detection \cite{ChangeDetection13}, scene recognition and view detection \cite{SUN360}, floor-plan construction \cite{Floorplan14} etc. These applications utilize the content in a panorama without image editing or modification. In terms of panoramic image editing, Schr\"{o}der and Sweldens \cite{Wavelets95} addressed the problem of sharpening and smoothing 360$^\circ$ panoramas via spherical wavelets. B\"{u}low \cite{Harmonic04} used spherical harmonic functions to smooth 3D surfaces, which can be regarded as panoramas. Kazhdan and Hoppe~\cite{Metric_2010} proposed a panorama enhancement method based on a metric-aware solver. Zhu et al. \cite{PCfSV2015} proposed a panorama completion method, in which the hole region is filled after it has been warped to 2D plane.

In this paper, we address the problem of image cloning for 360$^\circ$ panoramas, which clones a patch of the source panorama to the target panorama.
Traditional methods for planar image cloning typically solve a Poisson equation with Dirichlet boundary condition defined by the target image~\cite{PoissonEditing03,DragDrop06}. The fundamental machinery is to construct a Laplacian membrane that smoothly interpolates the difference between the source and target images along the boundary across the entire cloned region. Later on, an approximation approach based on mean value coordinates was originally proposed in \cite{MVC_Cloning_2009}. Because of its advantages in terms of speed, memory footprint, and parallelizability, the coordinate-based method has been extended in several works \cite{Harmonic10,Intent13,Chen2013}.

Considering 360$^\circ$ panorama cloning, an intuitive user interface is to perform the cloning over the latitude-longitude planar format. However, planar image cloning methods are not directly applicable. One major difficulty lies in that we should guarantee the cloned patch to have consistent geometric deformations with the target image wherever the cloned patch is pasted.
It is noted that panorama is subjective to the sphere geometry constraint.
Hence, we treat the cloned patch as a spherical polygon, and extend the coordinate-based cloning technique to the spherical domain.  %Specifically, we extend the coordinate-based cloning technique to seamlessly panorama cloning with spherical mean value coordinates as intensity differences diffusion weights.
Our work has three contributions:
% description
% enumerate
% itemize
\begin{enumerate}
  \item We are the first to study 360$^\circ$ panorama cloning, a new problem belonging to panorama editing, and get pleasing cloning results.
%  \item We re-derive the spherical mean value coordinates using stereographic projection, which is helpful for identifying discoloration problems in the large patch cloning.
  \item We explain how to preserve the orientation of cloned patches, which is an inherent problem of panorama cloning, with a two-step rotation estimation method.
  \item We remove discoloration artifacts when cloned patches cover over 180$^\circ$ filed-of-view by a splitting-based method, which is based on our new deduction of spherical mean value coordinates.

%We propose a two-steps rotation computation method to preserve the cloning patch's orientation, and a splitting method to remove the discoloration artifacts when cloning large patches, i.e. the patches covering more than 180$^\circ$.
\end{enumerate}

\section{Related Work}\label{sec:Work}

\subsection{Panorama-based Applications}
By providing wide fields of view far beyond planar images, 360$^\circ$ panoramas have been successfully applied in a number of recent new applications. For instance, online street-level virtual navigation~\cite{StreetView,Cube2Video13} synthesizes new panoramas at intermediate views in-between pre-captured ones. City-scale change detection~\cite{ChangeDetection13} estimates architectural structure changes between a panorama collection and a maintained cadastral 3D model. Xiao et al. \cite{SUN360} proposed to recognize image scene and view direction based on a large panorama dataset. Micusik and Kosecka \cite{Reconstruction} constructed 3D city model from street-view panoramic sequences. There also exist several works on panorama smoothing and sharpening, using spherical wavelets \cite{Wavelets95}, spherical harmonics functions \cite{Harmonic04}, and metric-aware solver \cite{Metric_2010}. Zhu et al. proposed a panorama completion method for street view images \cite{PCfSV2015}, which needs to warp the hole region to 2D plane first. In this paper, we focus on panorama cloning, a new problem that has not been studied yet to the best of our knowledge.

%\if lwan-0
%by comparing a panorama sequence to a historically collected one
%
%Schr\"{o}der and Sweldens demonstrated smoothing and sharpening of environment maps using spherical wavelets. Kazhdan and Hoppe~\cite{Metric_2010} proposed metric-aware solver for Laplacian processing of spherical data, which can be applied to enhance the spherical images. As the surface (mesh) of a 3D object can be represented as a function on the sphere, some surface (mesh) editing algorithms can be used for spherical image editing, such as surface smoothing based on spherical harmonic functions
%\fi

\subsection{Image Cloning}
Traditional image cloning methods \cite{PoissonEditing03,DragDrop06,MPB2015} eventually solve a large sparse linear system defined by the Poisson equation. Solving the Poisson equation for large cloned patches is a computing and memory intensive task. Although some acceleration algorithms~\cite{GPUPoisson} were proposed, they are still time consuming for real-time applications.

Coordinate-based techniques \cite{MVC_Cloning_2009,Harmonic10} offer practical alternatives for image cloning. These methods diffuse the differences along the patch's boundary to the interior region by interpolation without solving Poisson equations. The widely used diffusion weights are mean value coordinates \cite{MVC_Cloning_2009} and harmonic coordinates \cite{Harmonic10}. Coordinate-based methods have nice properties, including high speed, small memory footprint and ease of parallelization. They have recently been extended to video composition \cite{Chen2013} and stereoscopic image cloning \cite{SteroCloning}. Our panorama cloning belongs to the coordinate-based technique.

\subsection{Mean Value Coordinates}
The mean value coordinate~\cite{Floater03} motivated by mean value theorem has important properties, such as smoothness, linear independence, and refinability. It can be used to approximate a harmonic-like solution to the boundary interpolation problem, and has many applications in computer graphics~\cite{MVC_ToG_2006}, such as Phong shading for arbitrary polygons, image warping and image cloning~\cite{MVC_Cloning_2009}.

Many extensions have been made to planar mean value coordinate. Ju et al. \cite{MVCfCTM2005} generalized mean value coordinates from closed 2D polygons to closed triangular meshes, which are continuous and smooth on the interior of meshes. Langer et al.~\cite{SMVC_2006} proposed \emph{spherical mean value coordinate} (\emph{SMVC}), which gives the coordinates for point $\mathbf{v}$ located on the unit sphere with respect to a spherical polygon. By replacing the linear precision property of planar coordinates by a requirement in terms of center of mass, \cite{BCoS2010} introduced a method for defining and computing barycentric coordinates with respect to polygons on general surfaces. Panozzo et al. \cite{WAoS2013} presented an efficient method for computing and inverting weighted averages on surfaces for point on surfaces. Our panorama cloning uses spherical mean value coordinates as diffusion weights because of the embedded sphere geometry.

\section{Overview of Spherical Image Cloning}\label{sec:Basic}

\begin{figure}[!t]
\centering
\includegraphics[width=0.45\linewidth]{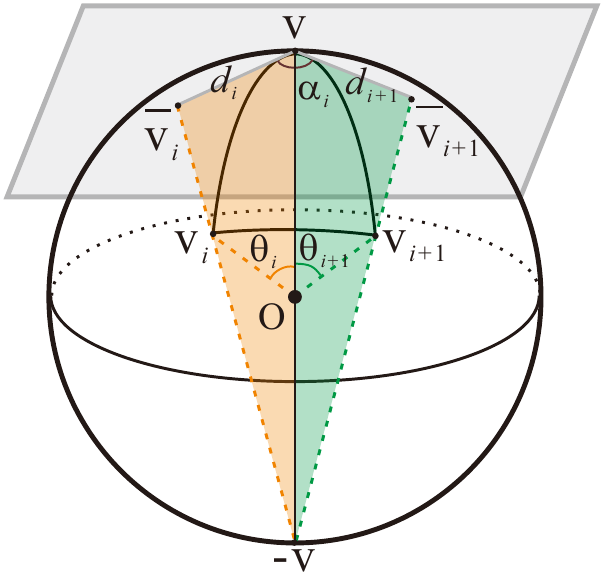}
\caption{Notation for the deduction of the spherical mean value coordinates, see Appendix A for more details.}
\label{fig:SMVC}
\end{figure}

Let $S=\mathds{S}^2$, $T=\mathds{S}^2$ be the domain of source and target panoramas, $g:S\rightarrow R, f^*:T\rightarrow R$ be the intensities of source and target image. Assume that we want to seamlessly clone the source patch $\Omega_{s}\subset S$ to the target patch $\Omega_{t}\subset T$. Similar to the planar image cloning presented by Farbman et al. \cite{MVC_Cloning_2009}, the intensity of the point $\mathbf{v}\in \Omega_{t}$ will be given by
\begin{equation}\label{equ:cloning}
f(\mathbf{v})=g(\mathbf{v})+\sum_{i=1}^{n}\lambda_{i}(\mathbf{v})(f^*-g)(\mathbf{v}_{i}),
\end{equation}
where $\lambda_{i}(\mathbf{v})$ is the spherical mean value coordinates of $\mathbf{v}$ with respect to the boundary of patch $\Omega_{t}$, whose vertices are $\mathrm{P}=\{\mathbf{v}_{1},\ldots,\mathbf{v}_{n}\}$.
According to our deduction in Appendix~\ref{sec:deduction} and \cite{SMVC_2006} , $\lambda_{i}(\mathbf{v})$ is given by
\begin{equation}\label{equ:final}
\lambda_{i}(\mathbf{v})=\frac{(\tan\frac{\alpha_{i-1}}{2}+\tan\frac{\alpha_{i}}{2})/\sin\theta_{i}}{\sum\nolimits_{j}\cot\theta_{j}(\tan\frac{\alpha_{j-1}}{2}+\tan\frac{\alpha_{j}}{2})},
\end{equation}
%\begin{equation}\label{equ:final}\small
%\lambda_{i}(\mathbf{v})=\frac{\tan\frac{\alpha_{i-1}}{2}+\tan\frac{\alpha_{i}}{2}}{\sin\theta_{i}}/\sum\nolimits_{j}\cot\theta_{j}(\tan\frac{\alpha_{j-1}}{2}+\tan\frac{\alpha_{j}}{2}),
%\end{equation}
where $\theta_{i}$ is the angle between $\mathbf{v}$ and $\mathbf{v}_{i}$; $\alpha_{i}$ is the signed angle between vectors $\mathbf{v}\times\mathbf{v}_{i}$ and $\mathbf{v}\times\mathbf{v}_{i+1}$ (see Figure~\ref{fig:SMVC}).
Because the mean value interpolant is very smooth away from the boundary of the cloned region, we construct an adaptive mesh for the spherical polygon as in \cite{MVC_Cloning_2009} to avoid much of the computation.

It seems that spherical image cloning is quite similar to planar image cloning. However, we find that it suffers from two unique problems. First, we need to preserve the orientation of the cloned patch after transformation so as to avoid weird appearance. Second, cloning large patches may result in discoloration artifacts, which damage the appearance of the cloned patches seriously. In the next, we will discuss these two issues in details.

\section{Preserving Cloned Patch Orientation}\label{sec:orientation}

%As we noted, the coordinate based image cloning given by Equation \ref{equ:cloning} is a difference diffusion process.
Panoramas are usually captured with the horizon aligned with the equatorial line.
To compute the intensity difference $f^*-g$ in Equation~\ref{equ:cloning}, we need to determine the boundary of the target patch according to the source patch.
%The target patch is determined by the shape of source patch and specified target cloning position $\mathbf{v}_{t}$.
In planar image cloning, the target patch can be computed by translating the source patch with offset $\mathbf{v}_{s}-\mathbf{v}_{t}$, where $\mathbf{v}_{s}$ is the datum point, e.g. the centroid of the source patch, and $\mathbf{v}_{t}$ is the specified target cloning position. In spherical image cloning, since $\mathbf{v}_{s}$ and $\mathbf{v}_{t}$ are both on the surface of sphere, the transformation between $\mathbf{v}_{s}$ and $\mathbf{v}_{t}$ now becomes 3D rotation.

\begin{figure}[ht]
\centering
\begin{tabular}{cc}
\hspace{-0.2cm}\includegraphics[width=0.49\linewidth]{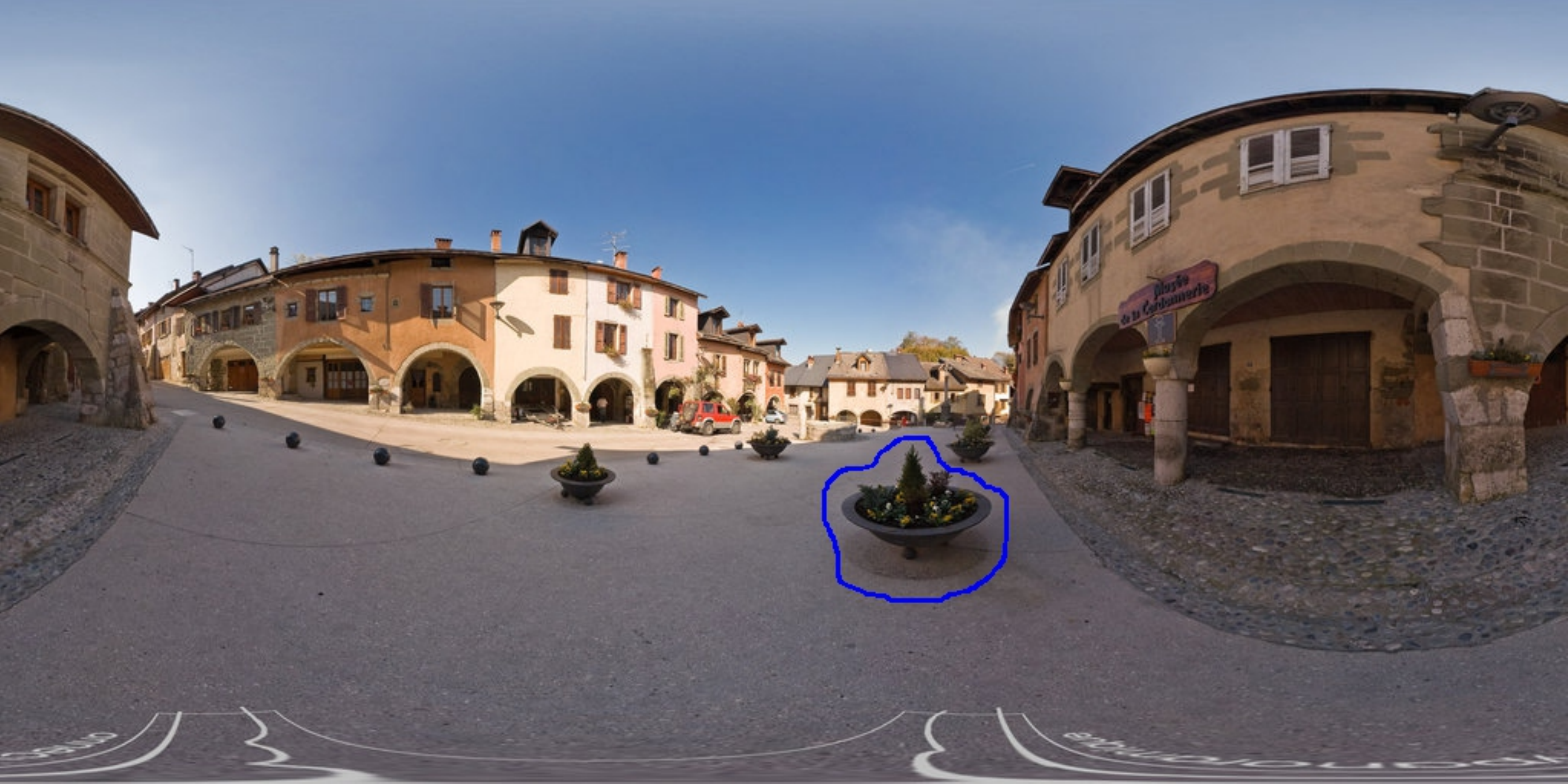}&
\hspace{-0.4cm} \includegraphics[width=0.49\linewidth]{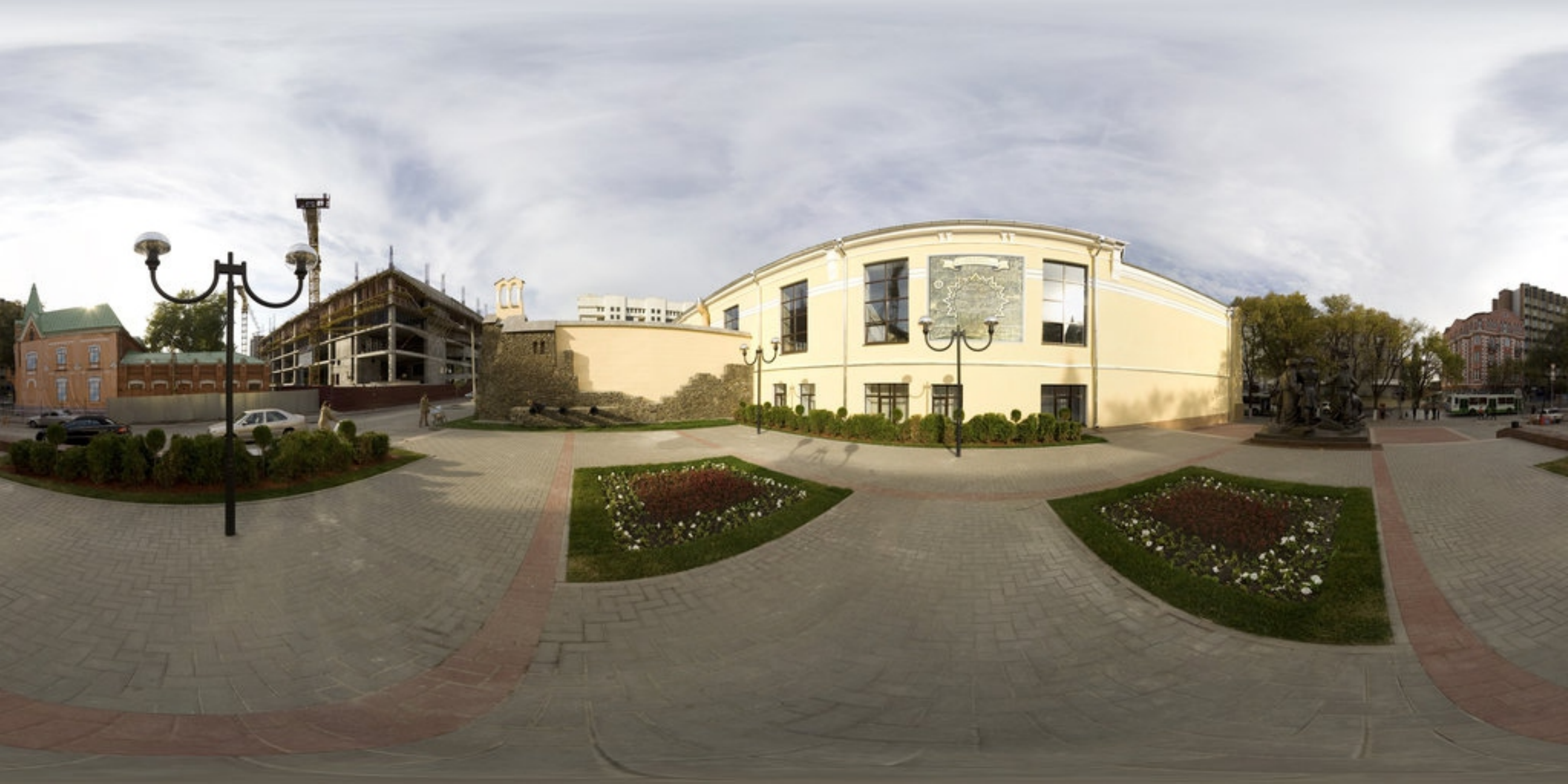}\\
\hspace{-0.2cm} (a)& \hspace{-0.4cm} (b)\\
\hspace{-0.2cm}\includegraphics[width=0.49\linewidth]{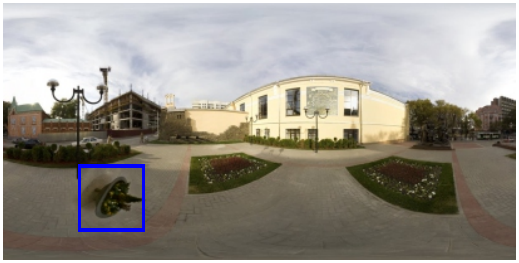}&
\hspace{-0.4cm} \includegraphics[width=0.49\linewidth]{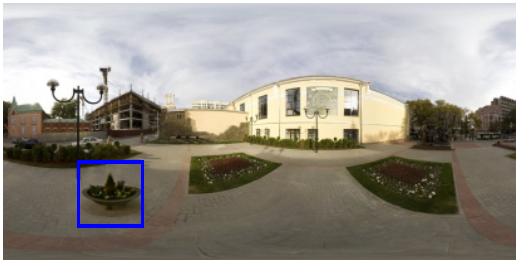}\\
\hspace{-0.2cm} (c)& \hspace{-0.4cm} (d)\\
\end{tabular}
\caption{The cloning results using different methods to compute the rotation matrix: (a) the source panorama and selected cloning patch; (b) the target panorama; (c) cloning result using naive method; (d) cloning result using two-step estimation method.}
\label{fig:orientation}
\end{figure}

%Given two points $\mathbf{v}_{s}$ and $\mathbf{v}_{t}$ on the sphere,
A naive method to compute the 3D rotation is using the axis-angle representation, which is given by Rodriguez' rotation formula
\begin{equation}
\mathcal{R}(\mathbf{u},\theta) = I\cos\theta+\sin\theta[\mathbf{u}]_{\times}+(1-\cos\theta)\mathbf{u}\otimes\mathbf{u},
\label{equ:rodriguez}
\end{equation}
where $\mathbf{u}=\mathbf{v}_{s} \times \mathbf{v}_{t}$ is the rotation axis and $\theta=\arccos(\mathbf{v}_{s} \cdot \mathbf{v}_{t})$ is the rotation angle. However this method may greatly change the orientation of cloned patch, leading to unexpected appearance as shown in Figure~\ref{fig:orientation}(c).
%after applying the transformation to the boundary vertices of source patch

\begin{figure}[!t]
    \centering
    \includegraphics[width=0.45\linewidth]{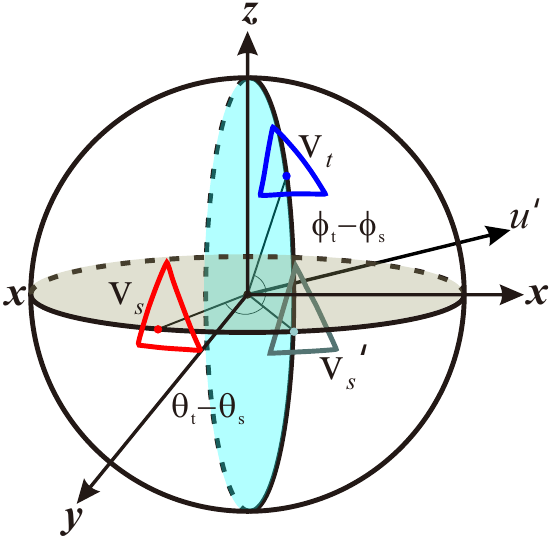}
\caption{To estimate the rotation between $\mathbf{v}_{s}$ and $\mathbf{v}_{t}$, we first rotate $\mathbf{v}_{s}$ to make it have the same azimuthal angle with $\mathbf{v}_{t}$, then rotate the intermediate point $\mathbf{v}_{s}{'}$ to $\mathbf{v}_{t}$.}
\label{fig:rotation}
\end{figure}

Here, we apply a two-step rotation estimation method to preserve the orientation of cloned patch (illustrated in Figure~\ref{fig:rotation}). Assume the spherical coordinates of $\mathbf{v}_{s}$ and $\mathbf{v}_{t}$ are $(\phi_{s}, \theta_{s})$ and $(\phi_{t}, \theta_{t})$ respectively, where $\phi_{s}$ and $\phi_{t}$ are azimuthal angles, $\theta_{s}$ and $\theta_{t}$ are polar angles. We first rotate $\mathbf{v}_{s}$ around $z$-axis by $\phi_{t}-\phi_{s}$ to make it have the same azimuthal angle with $\mathbf{v}_{t}$. This rotation is denoted as $R_{1}$, given by
\begin{equation}
R_{1}=
\left[
  \begin{array}{ccc}
    \cos(\phi_{t}-\phi_{s}) & -\sin(\phi_{t}-\phi_{s}) & 0\\
    \sin(\phi_{t}-\phi_{s}) & \cos(\phi_{t}-\phi_{s}) & 0\\
    0 & 0 & 1\\
  \end{array}
\right].
\end{equation}
The intermediate rotated point has the spherical coordinate $(\phi_{t}, \theta_{s})$ and is denoted as $\mathbf{v}_{s}{'}=R_{1}\mathbf{v}_{s}$.

The second step is to transform $\mathbf{v}_{s}{'}$ to $\mathbf{v}_{t}$.
%Because $\mathbf{v}_{s}{'}$ has the same azimuthal angle with $\mathbf{v}_{t}$,
This can be accomplished by rotating $\mathbf{v}_{s}{'}$ around the normal $\mathbf{u}{'}$ of blue-colored plane by $\theta_{t}-\theta_{s}$, where
\begin{equation}
\mathbf{u}{'}=\mathbf{v}_{s}{'}\times[0\ 0\ 1]^{T}.
\end{equation}
This rotation is represented by Rodriguez' rotation formula in Equation~\ref{equ:rodriguez} as
\begin{equation}
R_{2}=\mathcal{R}(\mathbf{u}{'},\theta_{t}-\theta_{s}).
\end{equation}
Hence, the final rotation $R$ between $\mathbf{v}_s$ and $\mathbf{v}_t$ is the combination of $R_{1}$ and $R_{2}$, given by
\begin{equation}
R=R_{2}R_{1}.
\end{equation}

The cloning results with rotation matrix computed by different methods are shown in Figure \ref{fig:orientation}. We can see that the orientation of the parterre is well preserved by the two-step rotation estimation method.

%\if 0
%Assume that the two points having spherical coordinates $(1,\theta_{s},\phi_{s})$ and $(1,\theta_{t},\phi_{t})$, where $\theta$ is the azimuthal angle and $\phi$ is the polar angle. $R_{1}$ is a rotation around $z$ axis and is given by
%\begin{equation}
%R_{1}=
%\left[
%  \begin{array}{ccc}
%    \cos\Delta\theta & -\sin\Delta\theta & 0\\
%    \sin\Delta\theta & \cos\Delta\theta & 0\\
%    0 & 0 & 1\\
%  \end{array}
%\right],
%\end{equation}
%where $\Delta\theta=\theta_{t}-\theta_{s}$ is the rotation angle. $R_{2}$ is a rotation around $\mathbf{u}=\mathbf{v}_{s}^{'}\times[0\ 0\ 1]^{T}$ and is given by
%\begin{equation}
%R_{2} = I\cos\Delta\phi+\sin\Delta\phi[\mathbf{u}]_{\times}+(1-\cos\Delta\phi)\mathbf{u}\otimes\mathbf{u},
%\end{equation}
%where $\Delta\phi=\phi_{t}-\phi_{s}$ is the rotation angle, $[\mathbf{u}]_{\times}$ is the cross product matrix of $\mathbf{u}$, $\otimes$ is the tensor product and $I$ is the identity matrix.
%\fi

%-------------------------------------------------------------------------
\section{Large Patch Cloning}\label{sec:Large}

\begin{figure}[ht]
\centering
\begin{tabular}{cc}
\hspace{-0.2cm}\includegraphics[width=0.49\linewidth]{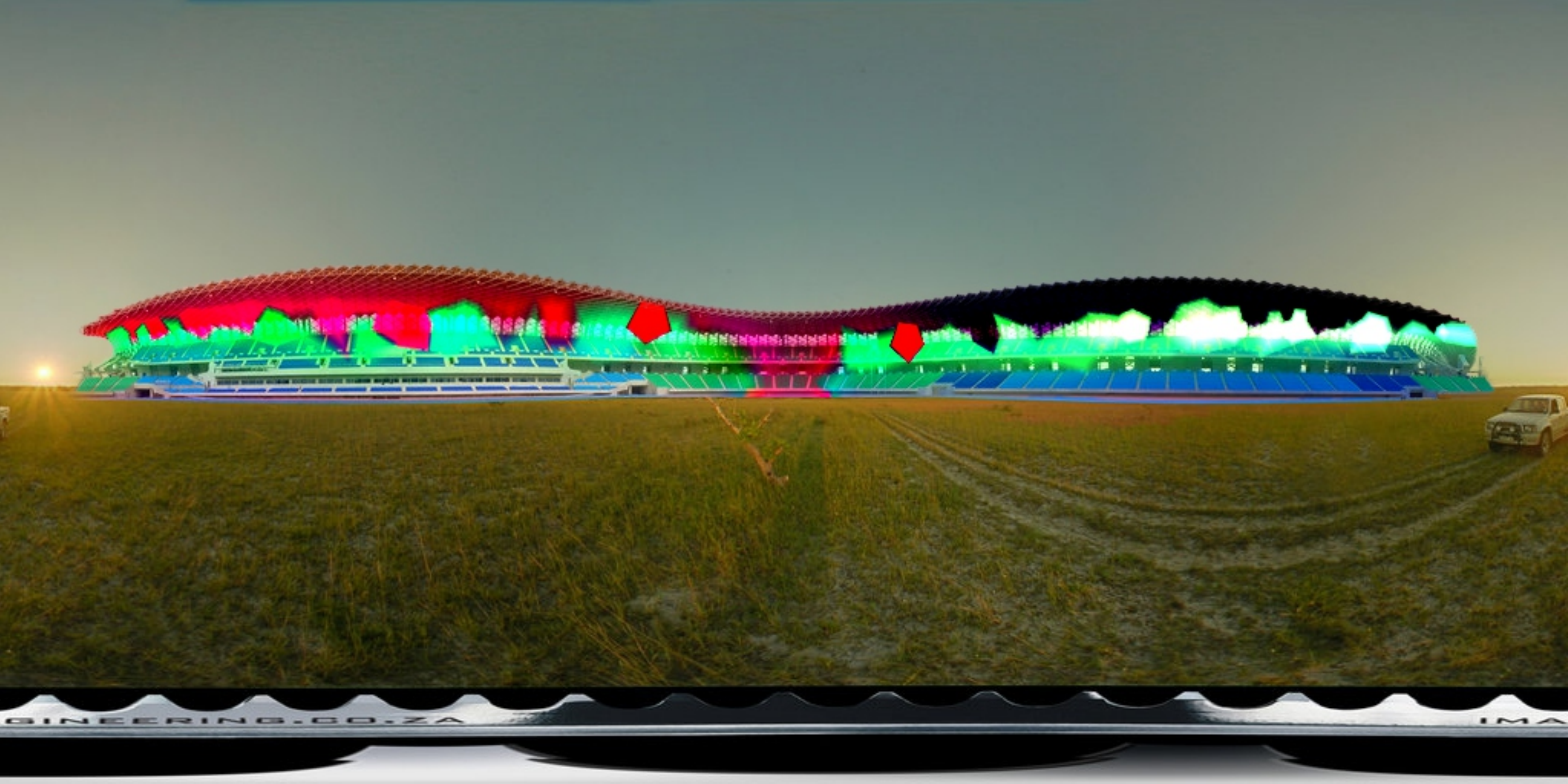} & \hspace{-0.3cm}\includegraphics[width=0.49\linewidth]{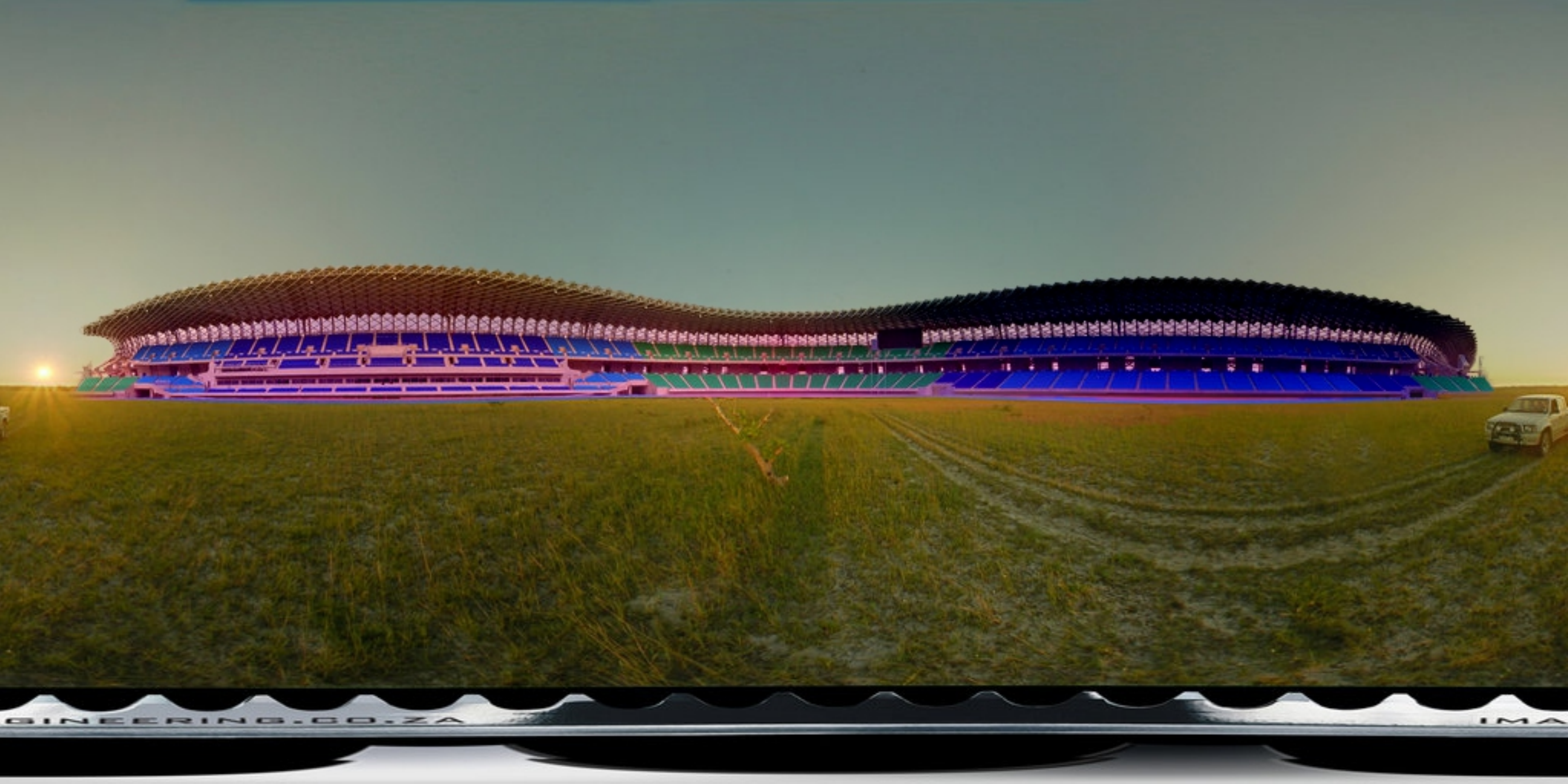}\\
\hspace{-0.2cm}
(a)& \hspace{-0.3cm} (b)\\
\end{tabular}
\caption{Discoloration artifacts. (a) Discoloration artifacts when the cloned patch covers more than 180$^\circ$. (b) The artifacts can be removed with splitting-based cloning.}
\label{fig:discoloration}
\end{figure}

The basic cloning method works well for small cloning patches, however it inevitably suffers from discoloration artifacts when we want to clone a quite large patch (see the example shown in Figure \ref{fig:discoloration}(a)). The main reason is that the spherical mean value coordinate given by Equation~\ref{equ:final} ``overflows'', when the angle $\theta_{i}$ between point $\mathbf{v}$ and boundary point $\mathbf{v}_{i}$ is more than 180$^\circ$.
The problem of discoloration artifacts is unique to spherical panorama cloning. Although some previous works \cite{Xie2010,Zhang2011,Chen2009,Chen2013} observed a similar problem, the artifacts they tackled are caused by large color differences or texture differences, which is not discussed in our paper.

\begin{figure*} [!t]
\centering
\includegraphics[width=0.98\linewidth]{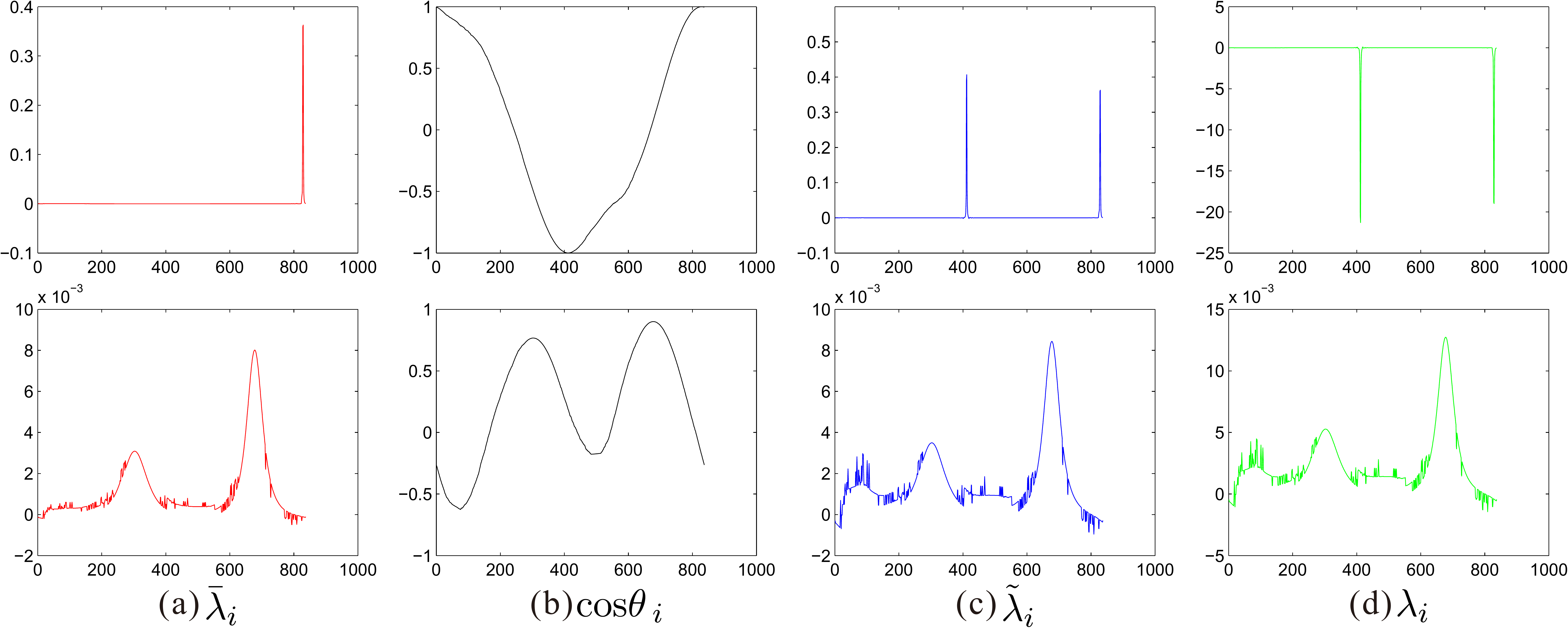}
\caption{Compared with the central pixels (the second row), the component of the spherical mean value coordinates for the pixels near the ends of the patch may have large absolute values. The horizontal axis represents the indices of pixels on the cloned patch boundary. See text for details.}
\label{fig:analysis}
\end{figure*}

\subsection{Analyzing Decoloration Artifacts}\label{sec:overfit}

From Figure~\ref{fig:discoloration} we notice that discolored pixels appear near two ends of the cloned patch, while central pixels have much less artifacts.
By further examining the discolored pixels, we find their spherical mean value coordinates (SMVC) have rather large magnitudes.
%This makes the diffused color differences overflowed, and hence causes the discoloration artifacts.
%To exploit the reason why these pixels have large magnitudes.
Although we currently have no way to prove this analytically, we try to give an explanation by examining the computation of SMVC.

To facilitate the analysis, we re-deduce SMVC by applying the stereographic projection in Appendix A. Here, we only recall critical relations.
Let $\bar{\lambda}_i$ be planar mean value coordinate of one point with respect to the projected polygon, $\tilde{\lambda}_i$ denotes intermediate coordinate, and $\lambda_i$ denotes the final spherical mean value coordinate with respect to the spherical polygon. From the deduction, the intermediate coordinate $\tilde{\lambda}_i$ is the scaling of $\bar{\lambda}_i$ (Equation \ref{equ:tildeLambda}), i.e.
\begin{equation}\label{eq:labmda1}
    \tilde{\lambda}_i=\frac{2}{1+\cos\theta_{i}}\bar{\lambda}_i.
\end{equation}
The spherical mean value coordinates $\lambda_i$ is related to $\tilde{\lambda}_i$ (Equation \ref{equ:lambda}) by
\begin{equation}\label{eq:labmda2}
\lambda_i=\frac{\tilde{\lambda}_{i}}{2-\sum_{i}\tilde{\lambda}_{i}}.
\end{equation}

Figure~\ref{fig:analysis} demonstrates the coordinate values for one central pixel and one discolored pixel (both from Figure~\ref{fig:discoloration}(a)). For the discolored pixel (the first row of Figure~\ref{fig:analysis}), there is one angle $\theta_i$ (between $\mathbf{v}$ and boundary point $\mathbf{v}_i$)  very close to $\pi$. Due to Equation \ref{eq:labmda1}, the corresponding $\tilde{\lambda}_i$ is abnormally enlarged as shown in Figure~\ref{fig:analysis}(c). In Equation \ref{eq:labmda2}, the sum $\sum_{i}\tilde{\lambda}_{i}$ would be close to 2, e.g. 2.0191 for this exampler discolored pixel. This makes $\lambda_i$ be negative and have large magnitude as shown in Figure~\ref{fig:analysis}(d). On the contrary, for the central pixel (the bottom row in Figure~\ref{fig:analysis}), the angle $\theta_i$ is far less than $\pi$.
%Then the scaling factor $\frac{2}{1+\cos\theta_{i}}$ would have relatively small absolute value. This makes
The curve of coordinate $\tilde{\lambda}_i$ has a similar shape with that of $\bar{\lambda}_i$.
%After we add $\tilde{\lambda}_{i}$s together, the sum $\sum_{i}\tilde{\lambda}_{i}$ is much smaller than 2, e.g. 1.3374 in this example.
As a result, $\lambda_i$ is the scaling of $\tilde{\lambda}_i$ with small factor, and has normal magnitude.
For the example in Figure~\ref{fig:discoloration}(a), the cloned region has a field of view around 300$^\circ$, and hence the discoloration artifacts seem to appear almost everywhere.

\subsection{Splitting-based Cloning}
\label{sec:splittingMethod}

%The decoloration problem is intuitively similar to the over fitting problem in polynomial curve fitting \cite{PRML06}. One technique that is often used to control the over fitting phenomenon in curve fitting is regularization, which adds a penalty term in order to discourage the coefficients of polynomial from reaching large values. However such technique can not be applied in our case. The main reason is that the curve fitting is always a minimization problem, while mean value coordinate involves an equality $\sum\nolimits_{i}\lambda_{i}\mathbf{v}_{i}=\mathbf{v}$.

Since no decoloration artifacts occur when $\theta_i$ (the angle between the interpolated point and boundary points) is less than 180$^\circ$, we think about the possibility to constrain the field of view of cloning patches. Encouraged by the fact that the central points in large cloning patches can get normal appearance, we develop a simple solution, by splitting the large cloning patch into two small patches, each of which has less than 180$^\circ$. In the following, we first derive the modified spherical mean value coordiantes.

%As long as each of the two resulted patches has a field-of-view smaller than $\pi$, the magnitudes of spherical mean value coordinates will be controlled effectively. By this way, the discoloration artifacts can be avoided. In the following, we explain the splitting-based cloning in more detail.

\begin{figure}[ht]
\centering
\includegraphics[width=0.75\linewidth]{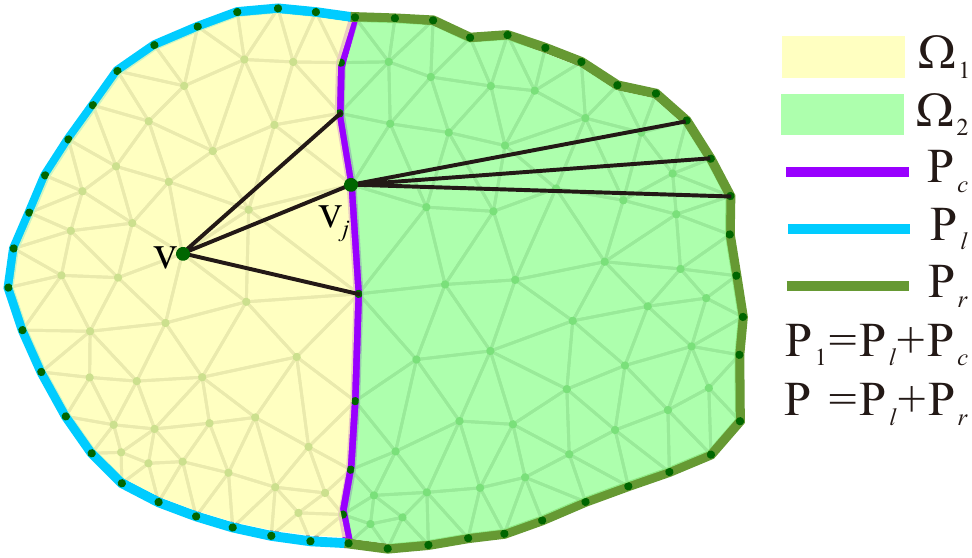}
\caption{The definition of sub regions and boundaries for splitting-based cloning.}
\label{fig:split}
\end{figure}

Suppose the adaptive mesh constructed in the cloned patch is $\Omega$, it is split into two sub-regions $\Omega_{1}$ and $\Omega_{2}$ along a path $\mathrm{P}_{c}$ (its selection will be discussed in Section \ref{sec:path}), as illustrated in Figure \ref{fig:split}.
Let $\mathrm{P}$ denote the boundary of $\Omega$, and $\mathrm{P}_{1}=\partial\Omega_{1}$, $\mathrm{P}_{2}=\partial\Omega_{2}$ denote the boundaries of the two sub-regions. Then the splitting path can be represented as $\mathrm{P}_{c}=\mathrm{P}_{1}\cap\mathrm{P}_{2}$. The vertices of the adaptive mesh are separated into two sets $A=\{\mathbf{v}|\mathbf{v}\in\Omega_{1}\}$ and $B=\{\mathbf{v}|\mathbf{v}\in\Omega_{2}\}$. For clarity, we introduce two auxiliary variables $\mathrm{P}_{l}=\mathrm{P}_{1}-\mathrm{P}_{2}$ and $\mathrm{P}_{r}=\mathrm{P}-\mathrm{P}_{l}$, which correspond to the sky blue line and the green line in Figure \ref{fig:split}.
For the vertices $\mathbf{v}\in A\cap B$, i.e. the vertices on the splitting path $\mathrm{P}_{c}$, their spherical mean value coordinates can be directly computed with respect to the original spherical polygon boundary $\mathrm{P}$, i.e.
\begin{equation}
\mathbf{v}=\sum\limits_{\mathbf{v}_{i}\in \mathrm{P}}\lambda_{\mathbf{v}\_\mathbf{v}_{i}}\mathbf{v}_{i},\ \mathrm{if}\ \mathbf{v}\in \mathrm{P}_{c},
\label{equ:v2P}
\end{equation}
where $\lambda_{\mathbf{v}\_\mathbf{v}_{i}}$ is the spherical mean value coordinate of $\mathbf{v}$ corresponding to $\mathbf{v}_{i}$.

For the vertices $\mathbf{v}\in A-B$, i.e. the rest vertices in $\Omega_{1}$, we first compute their spherical mean value coordinates with respect to $\mathrm{P}_{1}$ and get
\begin{equation}
\mathbf{v}=\sum\limits_{\mathbf{v}_{i}\in \mathrm{P}_{1}}\bar\lambda_{\mathbf{v}\_\mathbf{v}_{i}}\mathbf{v}_{i},\ \mathrm{if}\ \mathbf{v}\in A-B.
\label{equ:v2P1}
\end{equation}

\noindent Because $\mathrm{P}_{1}=\mathrm{P}_{c}\cup\mathrm{P}_{l}$,
% and $\mathrm{P}_{c}\cap\mathrm{P}_{l}=\emptyset$,
Equation~\ref{equ:v2P1} can be written as
\begin{equation}
\mathbf{v}=\sum\limits_{\mathbf{v}_{j}\in \mathrm{P}_{c}}\bar\lambda_{\mathbf{v}\_\mathbf{v}_{j}}\mathbf{v}_{j}+\sum\limits_{\mathbf{v}_{i}\in \mathrm{P}_{l}}\bar\lambda_{\mathbf{v}\_\mathbf{v}_{i}}\mathbf{v}_{i}.
\end{equation}
By substituting $\mathbf{v}_{j}\in \mathrm{P}_{c}$ with Equation~\ref{equ:v2P}, the above equation becomes
\begin{equation}\label{equ:vjc}
\begin{aligned}
\mathbf{v}=\sum\limits_{\mathbf{v}_{j}\in \mathrm{P}_{c}}\bar\lambda_{\mathbf{v}\_\mathbf{v}_{j}}\sum\limits_{\mathbf{v}_{i}\in \mathrm{P}}\lambda_{\mathbf{v}_{j}\_\mathbf{v}_{i}}\mathbf{v}_{i}+\sum\limits_{\mathbf{v}_{i}\in \mathrm{P}_{l}}\bar\lambda_{\mathbf{v}\_\mathbf{v}_{i}}\mathbf{v}_{i}\\
=\sum\limits_{\mathbf{v}_{i}\in \mathrm{P}}\sum\limits_{\mathbf{v}_{j}\in \mathrm{P}_{c}}\bar\lambda_{\mathbf{v}\_\mathbf{v}_{j}}\lambda_{\mathbf{v}_{j}\_\mathbf{v}_{i}}\mathbf{v}_{i}+\sum\limits_{\mathbf{v}_{i}\in \mathrm{P}_{l}}\bar\lambda_{\mathbf{v}\_\mathbf{v}_{i}}\mathbf{v}_{i}.
\end{aligned}
\end{equation}
Since $\mathrm{P}=\mathrm{P}_{l}\cup\mathrm{P}_{r}$, %and $\mathrm{P}_{l}\cap\mathrm{P}_{r}=\emptyset$,
we further rewrite Equation~\ref{equ:vjc} as
\begin{equation}
\begin{split}
\mathbf{v}=\sum\limits_{\mathbf{v}_{i}\in \mathrm{P}_{l}}\bar\lambda_{\mathbf{v}\_\mathbf{v}_{i}}\mathbf{v}_{i}+\sum\limits_{\mathbf{v}_{i}\in \mathrm{P}_{l}}\sum\limits_{\mathbf{v}_{j}\in \mathrm{P}_{c}}\bar\lambda_{\mathbf{v}\_\mathbf{v}_{j}}\lambda_{\mathbf{v}_{j}\_\mathbf{v}_{i}}\mathbf{v}_{i}\\
+\sum\limits_{\mathbf{v}_{i}\in \mathrm{P}_{r}}\sum\limits_{\mathbf{v}_{j}\in \mathrm{P}_{c}}\bar\lambda_{\mathbf{v}\_\mathbf{v}_{j}}\lambda_{\mathbf{v}_{j}\_\mathbf{v}_{i}}\mathbf{v}_{i}.
\end{split}
\end{equation}
Then the coordinates for vertices $\mathbf{v}\in A-B$ with respect to the original spherical polygon boundary $\mathrm{P}$ can be expressed as
\begin{equation}\label{equ:newlambda}
\lambda_{\mathbf{v}\_\mathbf{v}_{i}}=
  \left\{
   \begin{array}{lr}
\bar\lambda_{\mathbf{v}\_\mathbf{v}_{i}} + \sum\limits_{\mathbf{v}_{j}\in \mathrm{P}_{c}}\bar\lambda_{\mathbf{v}\_\mathbf{v}_{j}}\lambda_{\mathbf{v}_{j}\_\mathbf{v}_{i}},&\mathrm{if}\ \mathbf{v}_{i}\in \mathrm{P}_{l};  \\
\\
\sum\limits_{\mathbf{v}_{j}\in \mathrm{P}_{c}}\bar\lambda_{\mathbf{v}\_\mathbf{v}_{j}}\lambda_{\mathbf{v}_{j}\_\mathbf{v}_{i}}, &\mathrm{if}\ \mathbf{v}_{i}\in \mathrm{P}_{r}. \\
   \end{array}
   \right.
\end{equation}
%(Note that the $\lambda_{\mathbf{v}\_\mathbf{v}_{i}}$ on the right hand side of the equation is the spherical mean value coordinates of $\mathbf{v}$ with respect to $\mathrm{P}_{1}$, while the one on the left hand side is the coordinates of $\mathbf{v}$ with respect to $\mathrm{P}$.) And
These coordinates are used as the weights to diffuse intensity difference when cloning the patch. A similar way can be taken to compute the coordinates for the vertices $\mathbf{v}\in B-A$.

It is worthy pointing out that the coordinates computed by the splitting-based method are different from those computed by the basic method in Section~\ref{sec:Basic}.
%The magnitudes of all the coordinates in the right hand side of Eq. (\ref{equ:newlambda}) are in the rang (0,1), so the final value of $\lambda_{\mathbf{v}\_\mathbf{v}_{i}}$ is in (0,1).
%Although it is difficult to be proven analytically, we notice that
We think the reason is that when computing with respect to different polygons, the tangent and the reciprocal of the distance, both of which are nonlinear functions, will get different values.

%In the splitting-based method, we generally perform computation with three polygons, i.e. the original one and two smaller polygons due to the splitting. By constraining the major computation within smaller polygons, our method is able to circumvent the discoloration problem (as illustrated in Figure~\ref{fig:discoloration}).

%The splitting-based cloning can solve the over fitting problem due to the property that the coordinates computed using splitting-based method are not the same as those directly computed using Eq. (\ref{equ:final}). Although it is difficult to prove analytically, we try to explain it from two aspects. The first reason is that the tangent plane in the deduction of the spherical mean value coordinate is defined by vertice $\mathbf{v}$. Different $\mathbf{v}$ induces different tangent plane, which further gives different projected planar polygon $\bar{\mathrm{P}}$. The second and maybe more fundamental reason is that this property also holds true for planar mean value coordinates. It is due to the fact that when we compute the mean value coordinates with respect to different polygons, the tangent and the reciprocal of the distance, both of which are nonlinear functions, will get different values.

\subsection{Splitting Path}
\label{sec:path}

\begin{figure}[!t]
\centering
\includegraphics[width=0.5\linewidth]{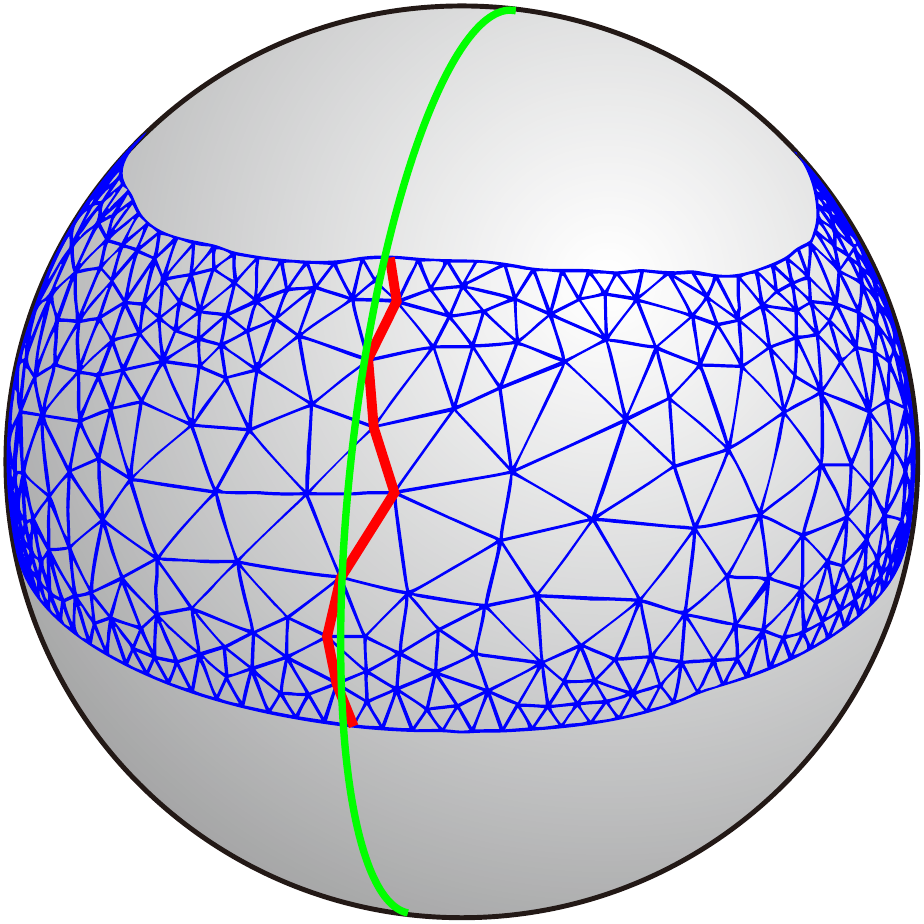}
\caption{Splitting path computation (see text for more details).}
\label{fig:path}
\end{figure}
The key point for the splitting-based cloning to work is breaking the large patch into two small patches, each covering less than 180$^\circ$.
%Before computing the coordinates, we should first find the splitting path along which the patch is splitted.
Because the source patches are selected on the unrolled spherical images, they unlikely cover more than 180$^\circ$ vertically.
Accordingly, a safe way to get two reasonable patches is vertically splitting the spherical polygon in the middle.

\begin{figure*}[!ht]
\centering
\begin{tabular}{ccc}
\includegraphics[width=0.48\linewidth]{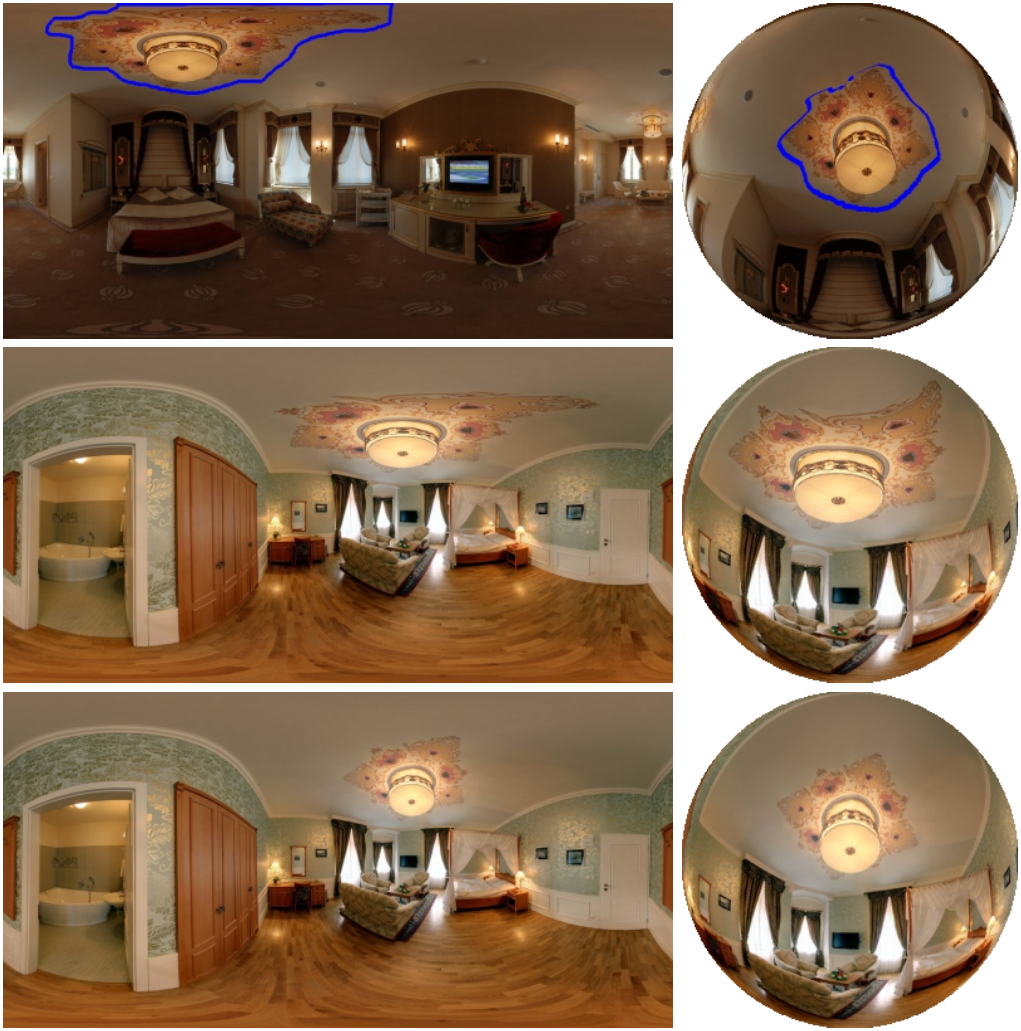}&\hspace{-0.2cm}\includegraphics[width=0.48\linewidth]{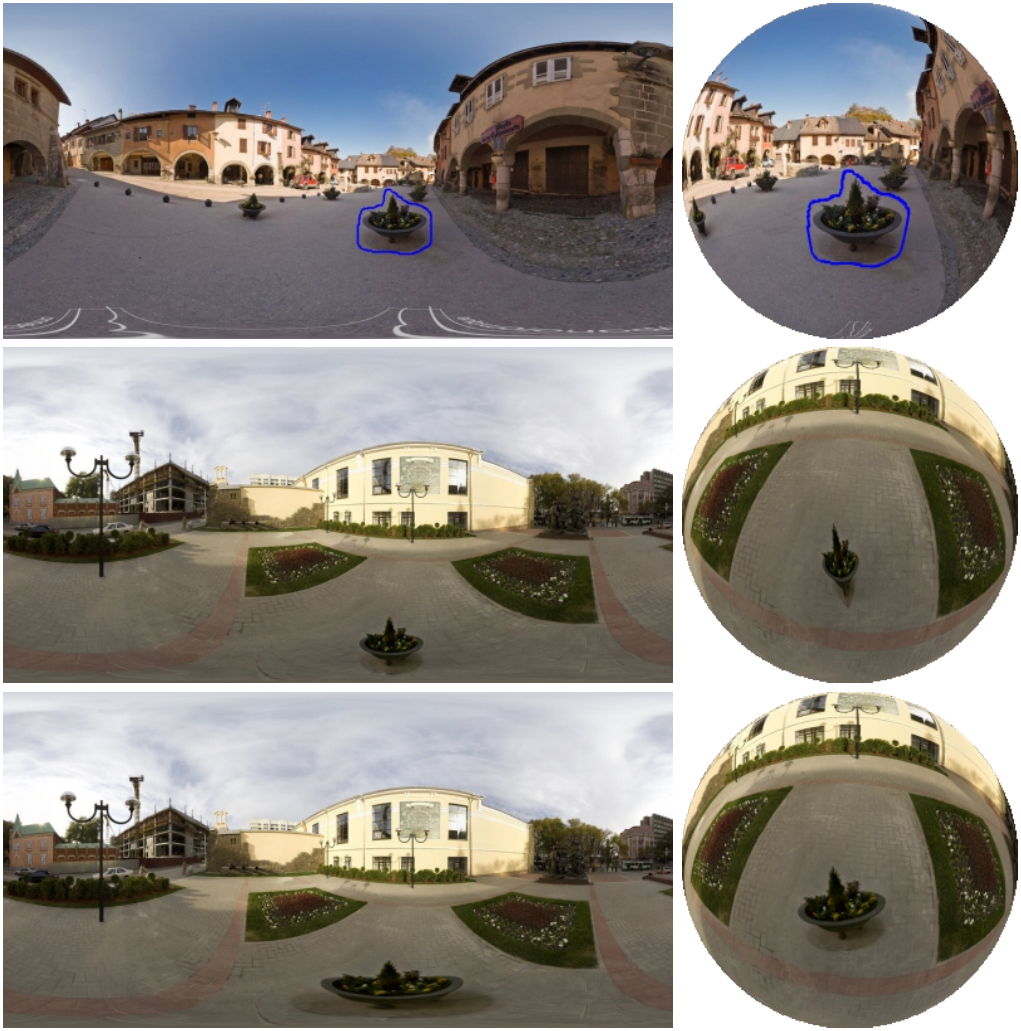} \\
(a)&(b)
\end{tabular}
\caption{Compared with the planar image cloning, our method can preserve the shape of the cloned object: (top) the source panorama and selected cloning regions, (middle) the cloning results using planar image cloning method \cite{MVC_Cloning_2009}, (below) the cloning results using our method.}
\label{fig:result1}
\end{figure*}

To be specific, we first find the median of the azimuthal angles of the boundary vertices. Based on it, a great circle  that passes through $z$-axis is constructed (the green curve in Figure \ref{fig:path}).
%The intersection points of this great circle and the spherical polygon is then determined.
%Next, we find the boundary vertices nearest to each of the intersection points, and mark them as the start node and the end node in the graph defined by the adaptive mesh. The splitting path (marked as the red curve in Figure \ref{fig:path}) is given as the shortest path between these two nodes.
We assume that the great circle and the spherical polygon have only two intersection points. Next, we find the boundary vertices nearest to each of the intersection points, and compute the splitting path (marked as the red curve in Figure \ref{fig:path}) as the shortest path of mesh points connecting the two boundary vertices. In addition to this splitting path generation method, we have tested another two solutions. As will be discussed in Section~\ref{sec:split}, our current method is easy to compute and able to generate consistently pleasing results.

\section{Experimental results}\label{sec:Result}

In the experiments, we use images from SUN360 panorama database~\cite{SUN360}. We first compare the cloning results of our method with those generated by planar image cloning, then we show the results generated with and without the splitting when the cloned region covers more than 180$^\circ$. After discussing the performance of the splitting-based cloning, we give our timing statistics.

\begin{figure*}[ht]
\centering
\begin{tabular}{ccc}
\includegraphics[width=0.48\linewidth]{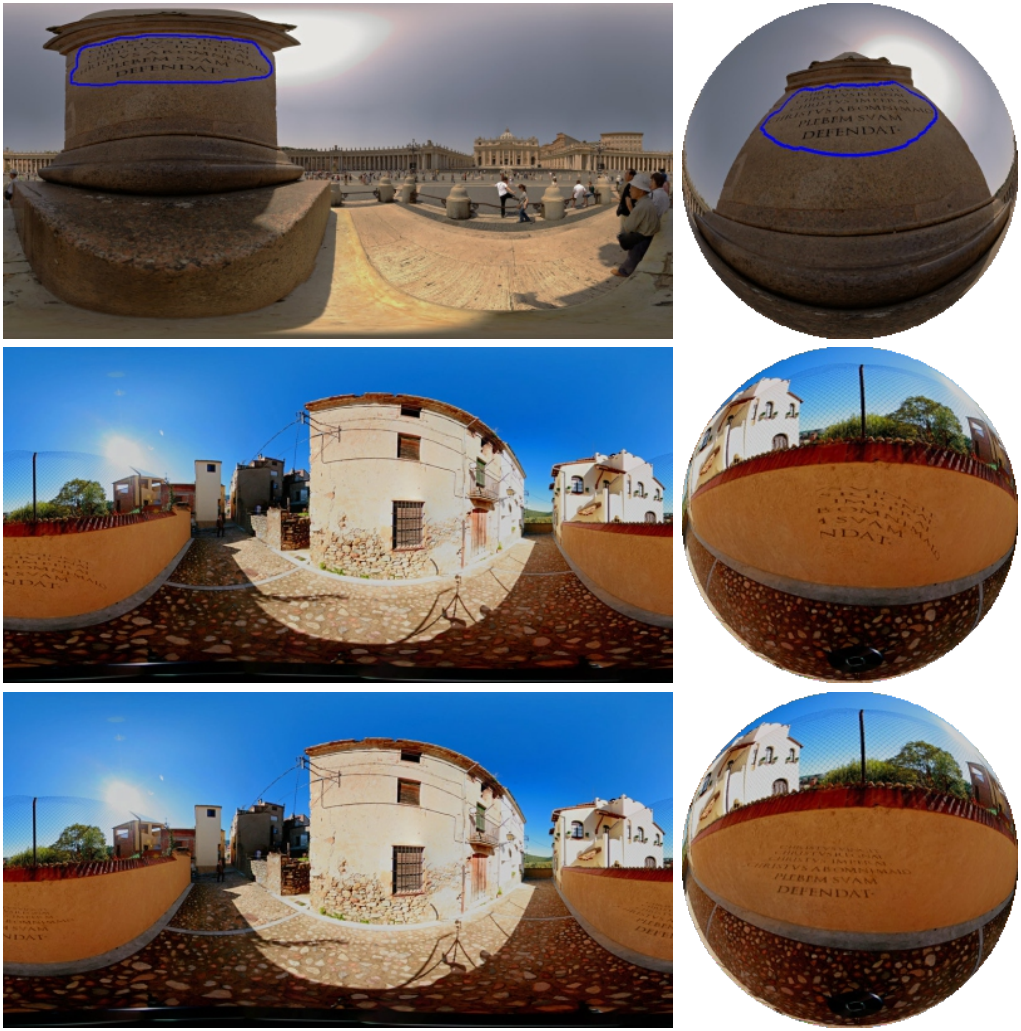}&\hspace{-0.2cm}\includegraphics[width=0.48\linewidth]{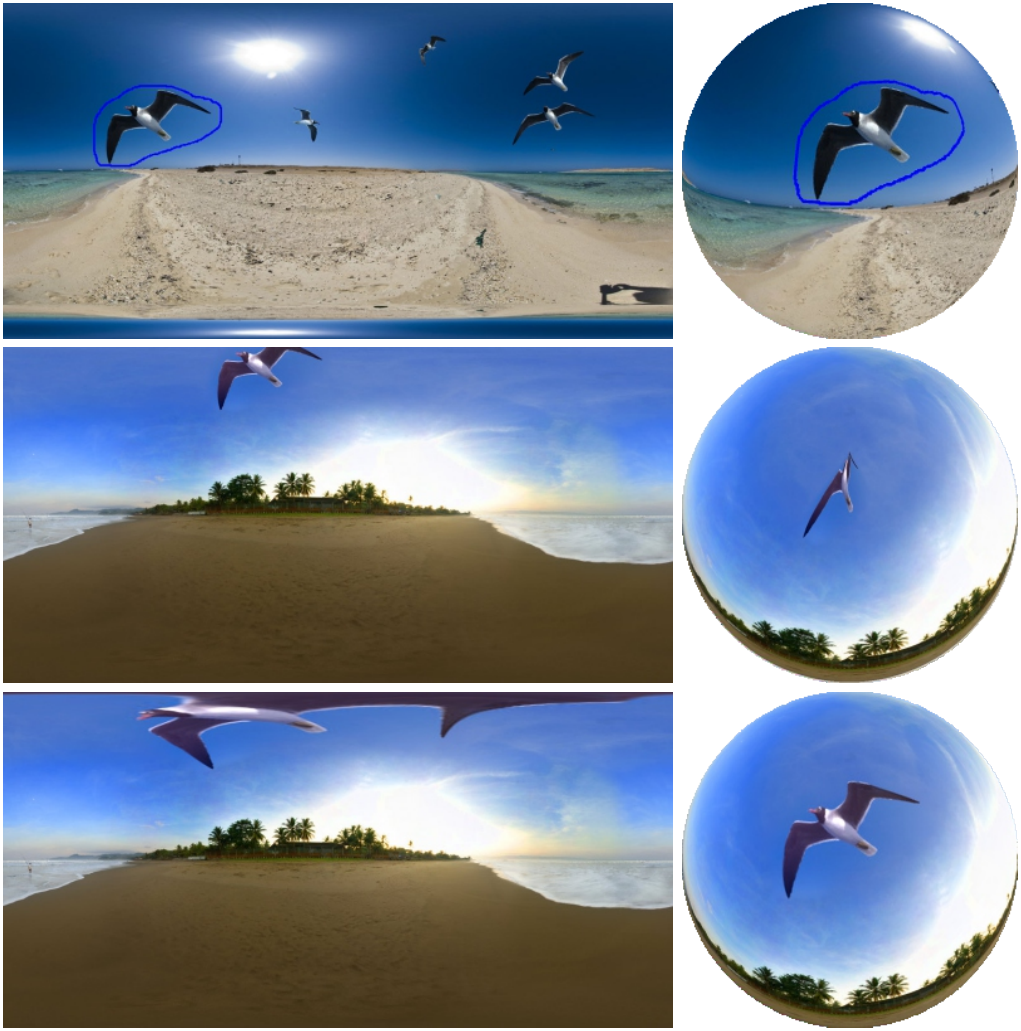} \\
(a)&(b)
\end{tabular}
\caption{Compared with the planar image cloning \cite{MVC_Cloning_2009}, our method does not suffer from the boundary problem when the cloning position gets close to the boundaries of the panorama.}
\label{fig:result2}
\end{figure*}

\subsection{Comparison with Planar Image Cloning}

Figure~\ref{fig:result1} shows the results by our method and the planar image cloning \cite{MVC_Cloning_2009}. Although it is hard to distinguish which results are better in the unfolded latitude-longitude panoramic format, it becomes clearer after mapping onto the sphere. In Figure~\ref{fig:result1} (a), we replace the lamp of a room with another one. In the result of the planar image cloning method, the original circular lamp becomes elliptical and the lamp decoration is stretched. On the contrast, our method preserves the lamp shape as well as the decoration. A similar performance is observed in Figure~\ref{fig:result1} (b), in which the flowerpot in the cloning result of the planar method is extruded, while our result does not suffer from the problem.

\begin{figure*}[ht]
\centering
\begin{tabular}{ccc}
\includegraphics[width=0.32\linewidth]{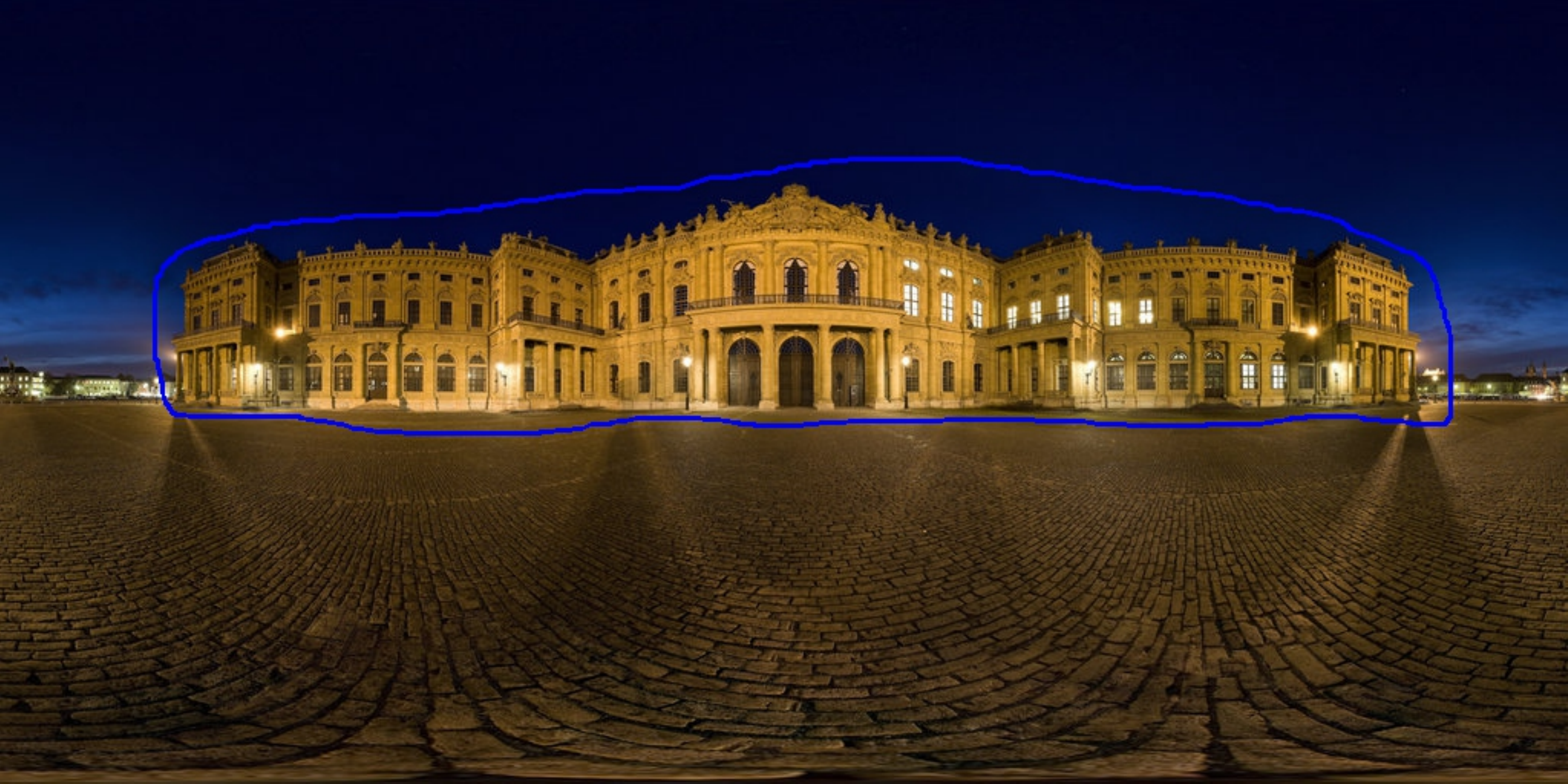}&\hspace{-0.3cm}\includegraphics[width=0.32\linewidth]{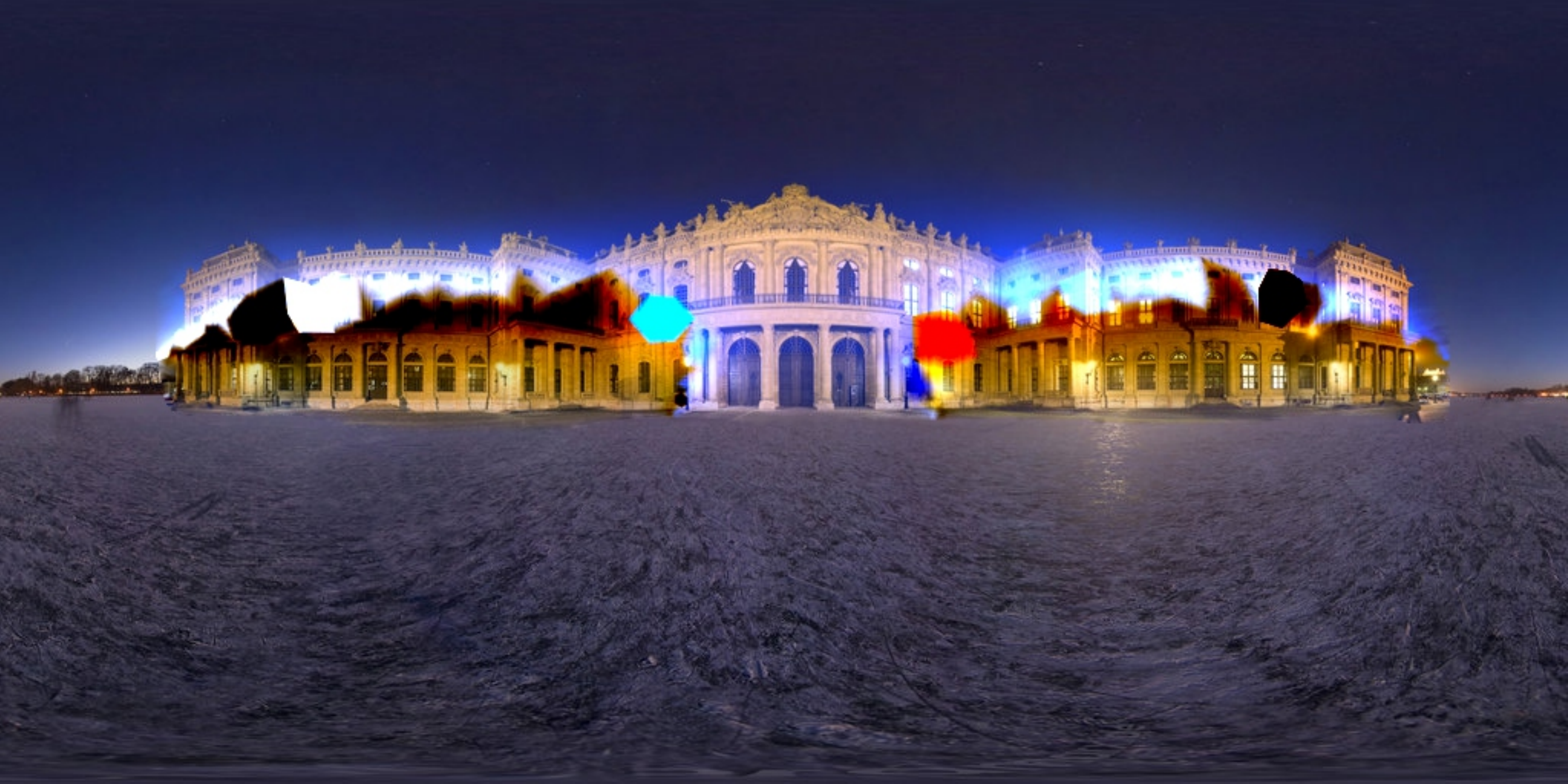} &\hspace{-0.3cm}\includegraphics[width=0.32\linewidth]{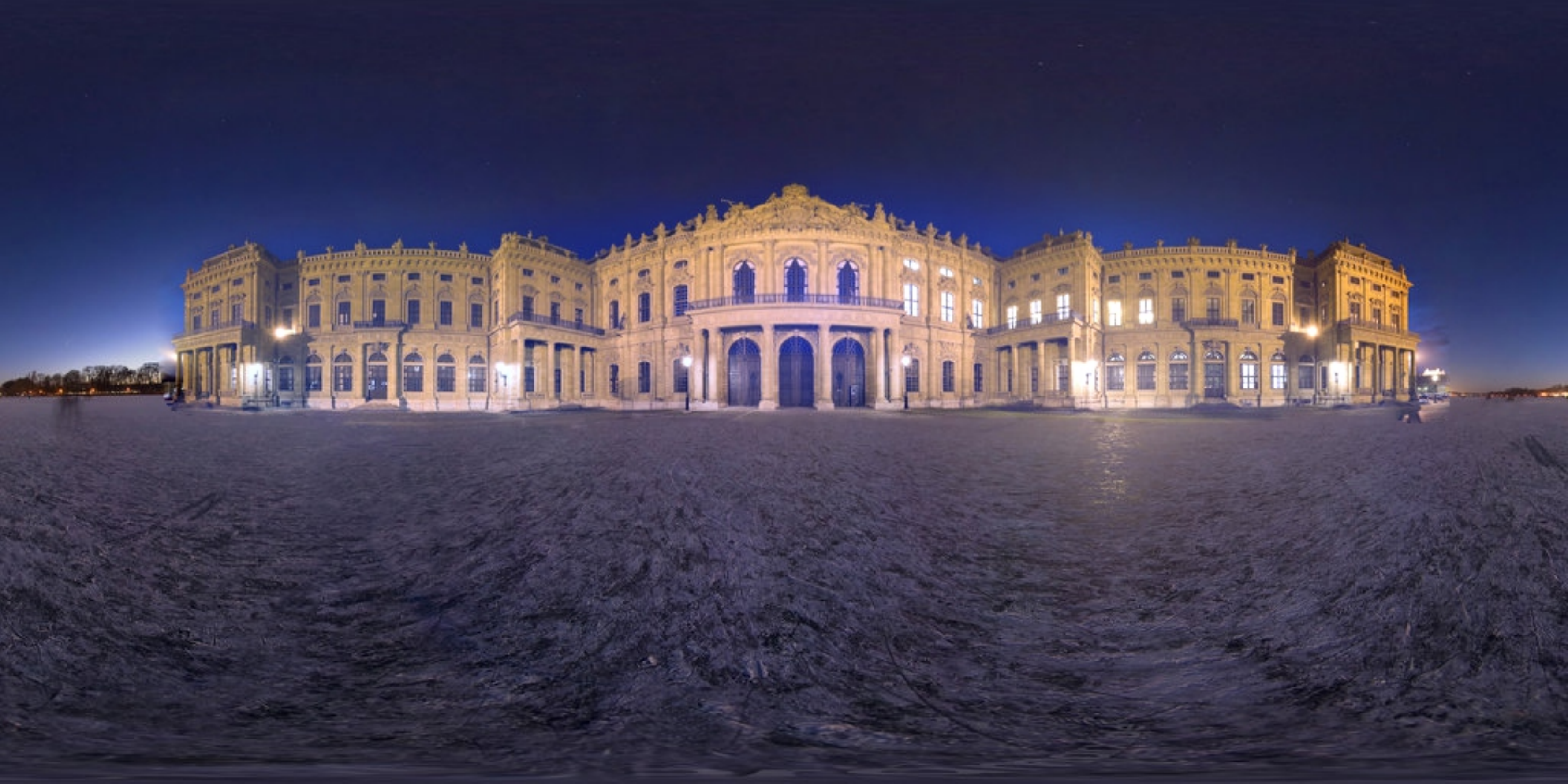}\\
\includegraphics[width=0.32\linewidth]{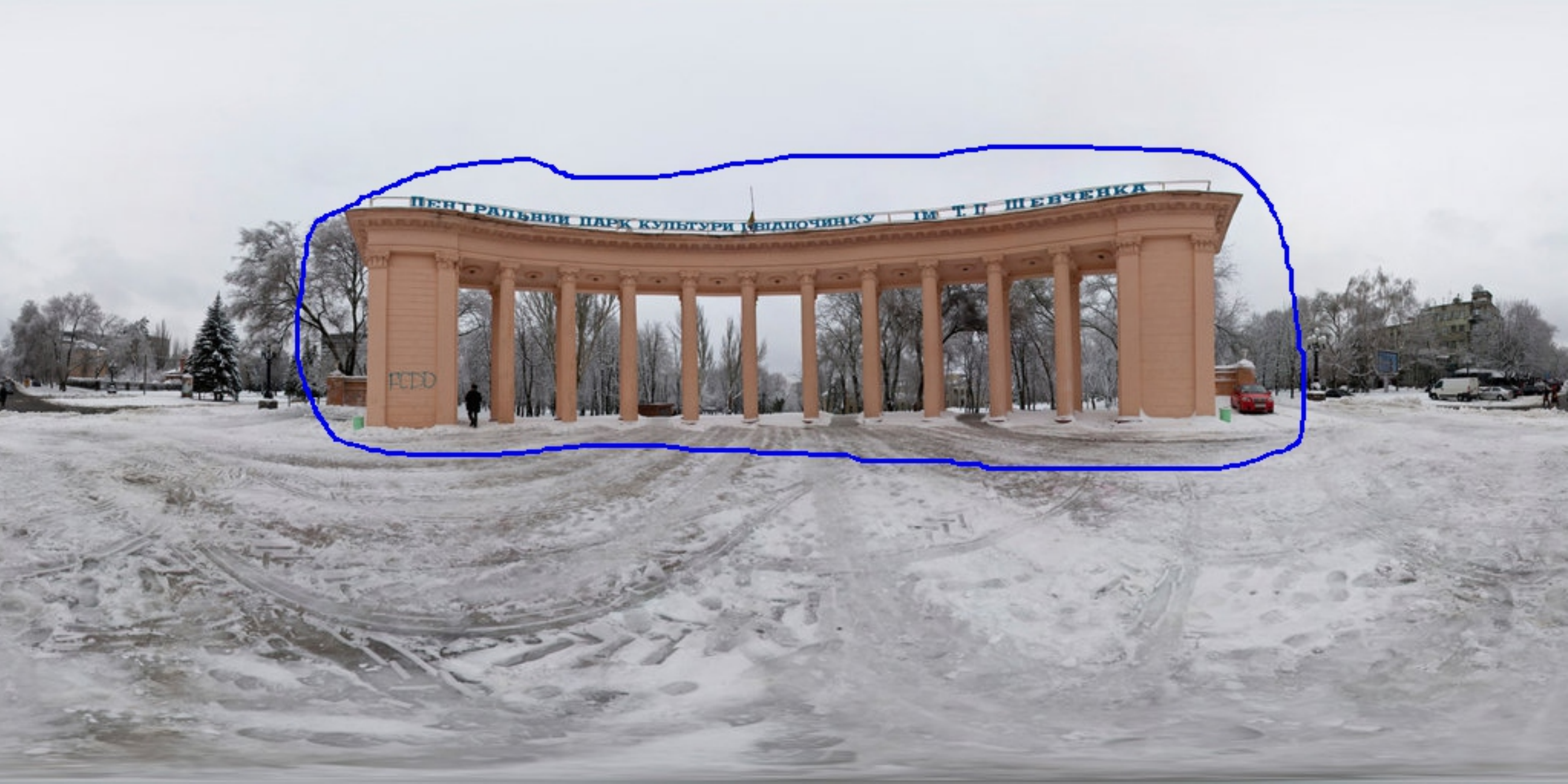}&\hspace{-0.3cm}\includegraphics[width=0.32\linewidth]{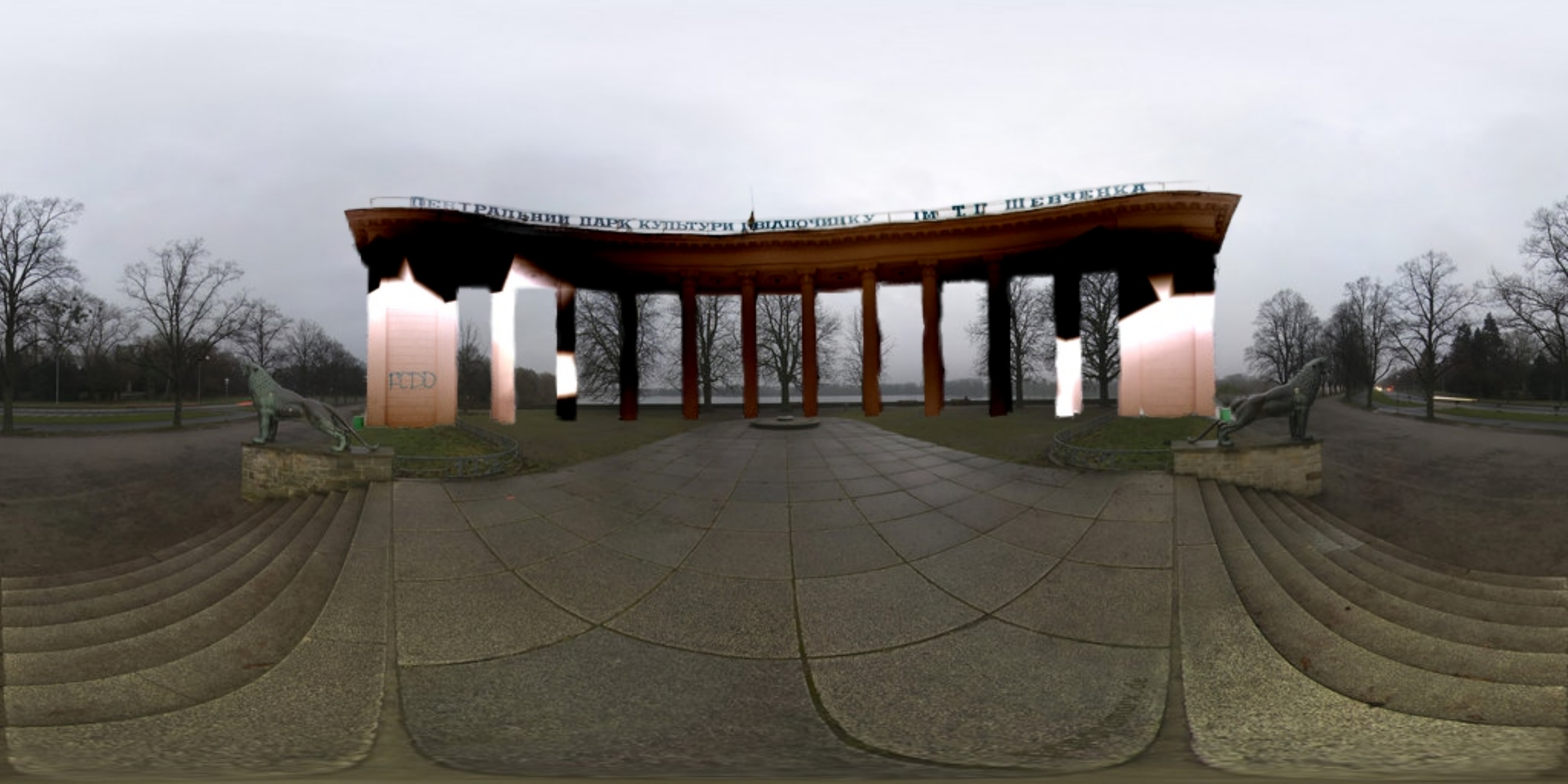} &\hspace{-0.3cm}\includegraphics[width=0.32\linewidth]{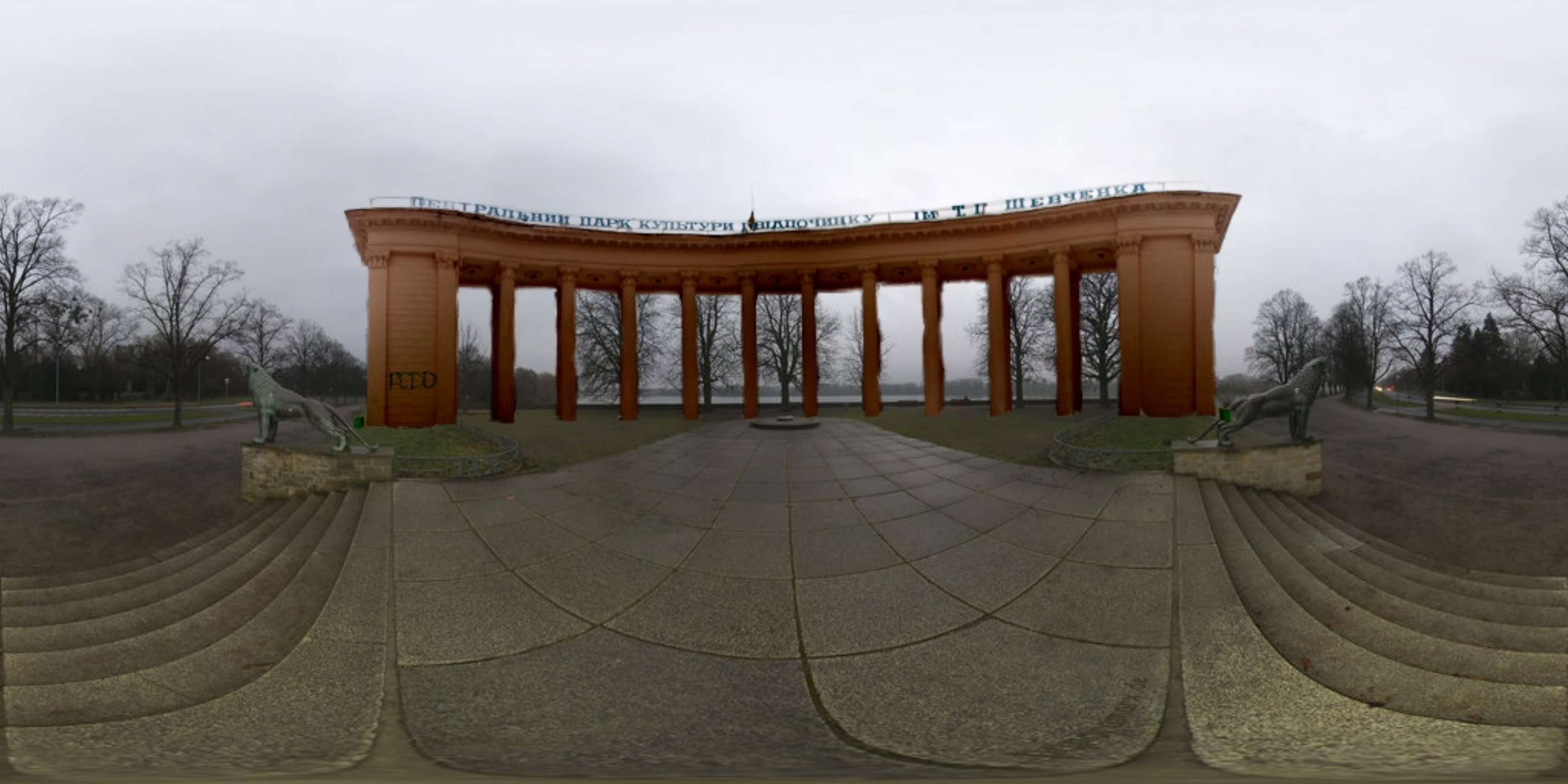}\\
\includegraphics[width=0.32\linewidth]{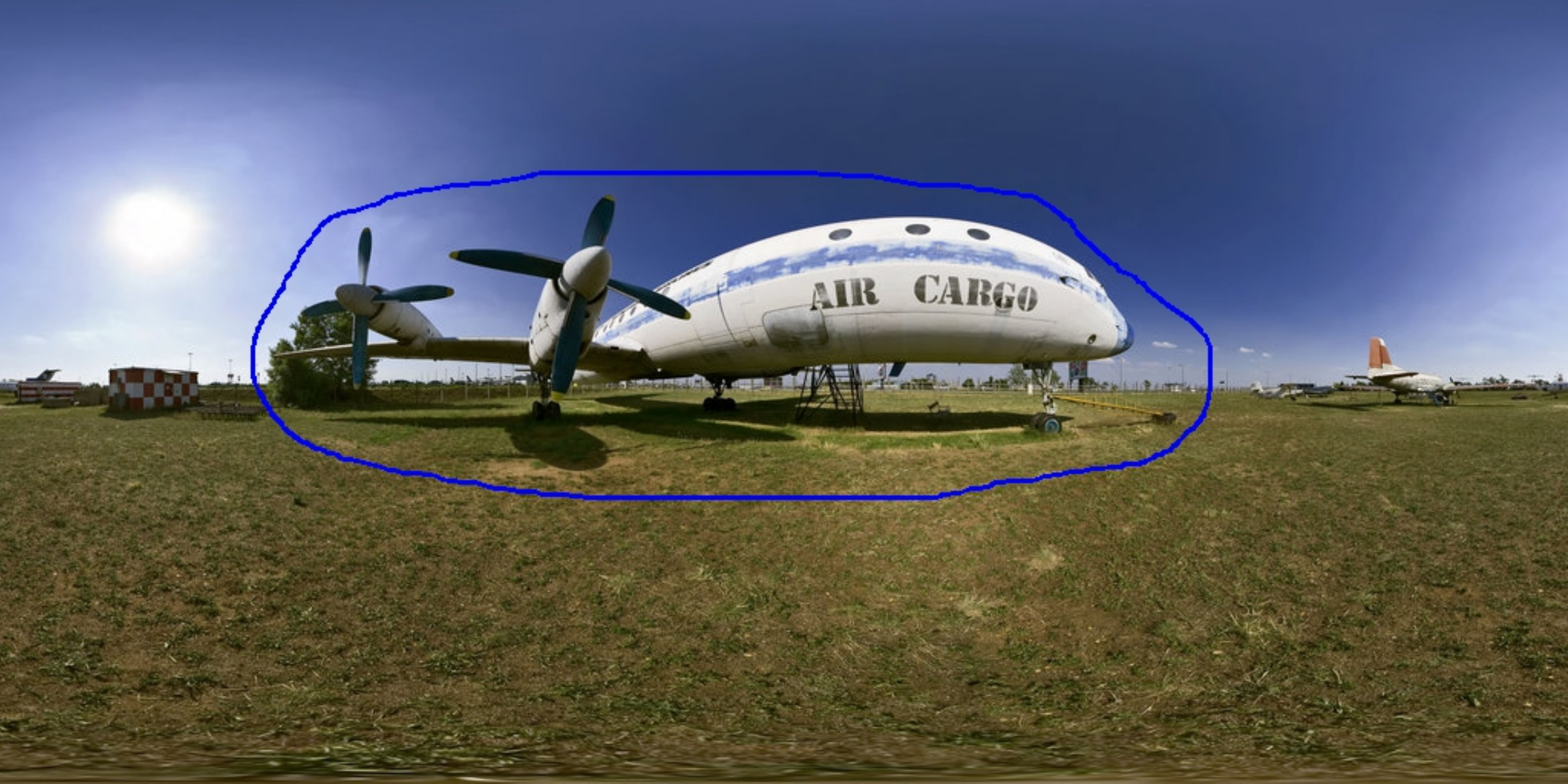}&\hspace{-0.3cm}\includegraphics[width=0.32\linewidth]{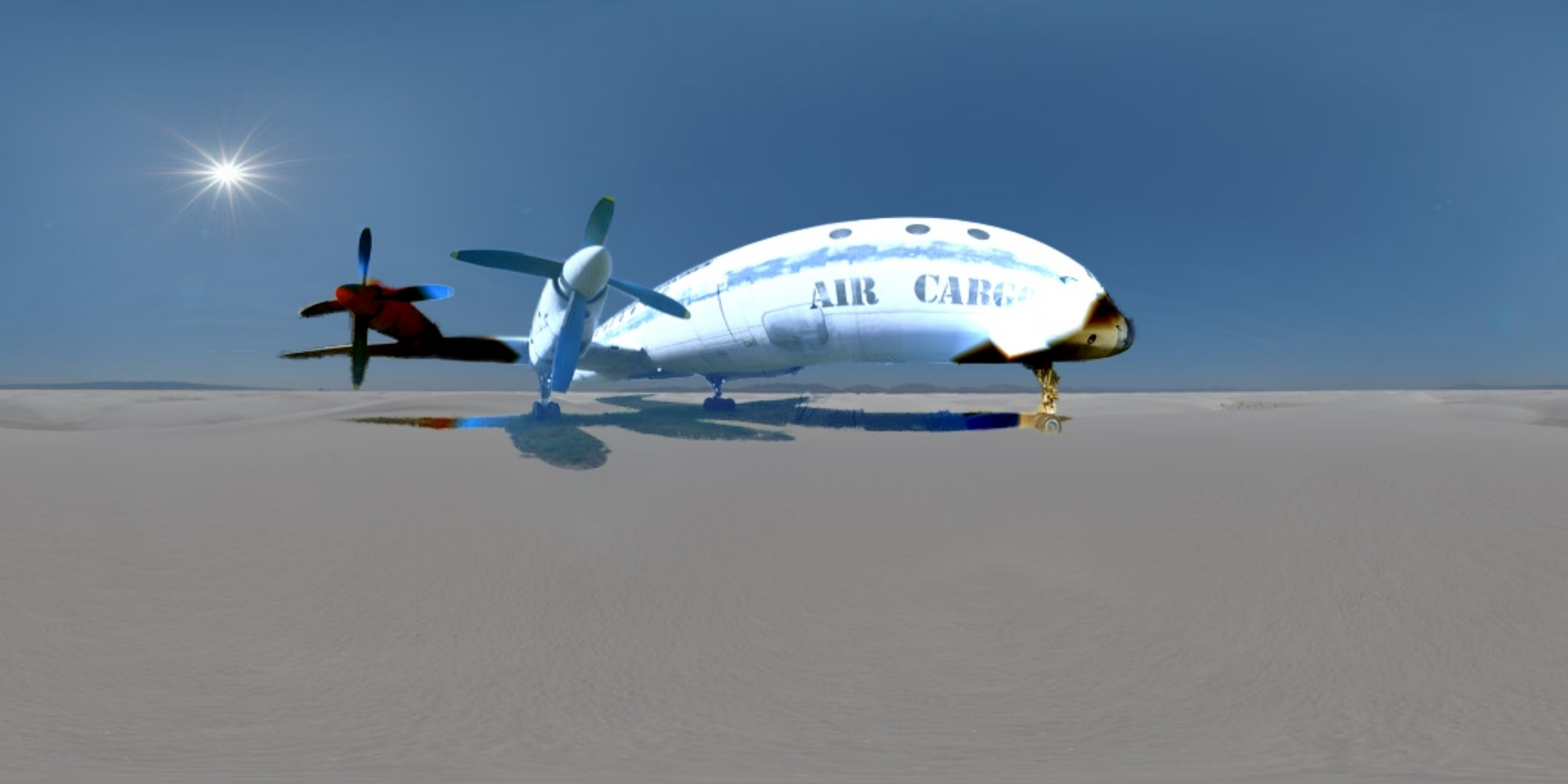} &\hspace{-0.3cm}\includegraphics[width=0.32\linewidth]{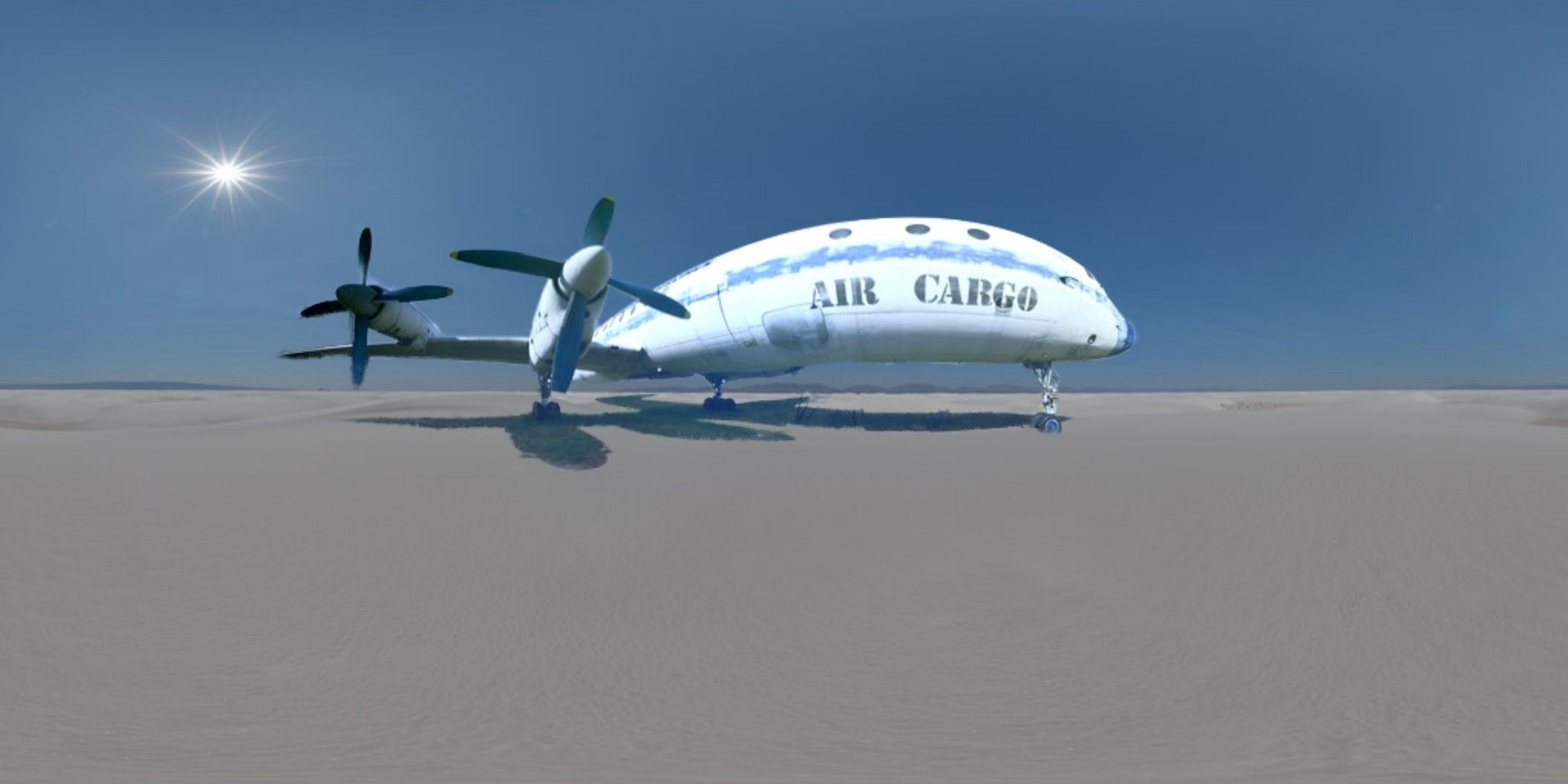}\\
(a)&\hspace{-0.3cm}(b)&\hspace{-0.3cm}(c)
\end{tabular}
\caption{When the cloned patch covers more than 180$^\circ$ (a), the cloning directly using spherical mean value coordinates will cause discoloration artifacts (b). Our splitting-based method can deal with this problem (c).}
\label{fig:result3}
\end{figure*}

Another advantage of our method is that it can naturally circumvent the boundary problem due to the spherical mapping. For the example in Figure \ref{fig:result2}(a), we clone the inscriptions on the source image to the wall of the target image. When the cloning position gets close to the left boundary of the target spherical image, the cloned region can automatically appear in the right part of the target image with our method. Although the same effect can be achieved by simple tricks in the planar image cloning, it can not deal with the case when the cloning position gets close to the top or bottom boundary of the target image. As shown in Figure \ref{fig:result2}(b), the planar image cloning results in a much distorted bird in the spherical domain. The main reason is that the left and right boundary of the panorama should be coincided and represent a meridian, while the top and the bottom boundary represent the north and south pole of the sphere respectively.

%\if 0
%From the mapped cloning results, we can see that our method can make the cloned patch has consistent deformations with the target image and keep the shape of the cloned objects unchanged.
%\fi

\subsection{Large Patch Cloning}
Figure \ref{fig:result3} gives some examples of cloned regions covering more than 180$^\circ$ field-of-view. On the first row, we clone a long building between two images. For the result generated by the panorama cloning without splitting, the two wings of the cloned building are discolored and darker than the main part. This is because the target image has higher intensities than the source image, and after we diffuse positive intensity differences with negative weights whose magnitudes are large (see Section \ref{sec:splittingMethod}), the intensity becomes very small. For the result generated by the cloning with splitting, the intensity of the cloned region is more smooth, and gives more pleasant appearance.

The second row gives another example, in which the building in the source image is placed on the platform in the target image. Contrasted with the first example, the two ends of this building are brighter than the central part if cloning without splitting. This is due to the fact that the intensity of the target image in this example is lower than that of the source image. As the cloned region is more complicated, if we directly clone the selected patch to the target image, the cloned region will conflict with the surrounding background of the target image. To make them match well, we compute a matte from the source patch, and remove the unwanted part by modulating the cloned region with the matte. The third row in Figure \ref{fig:result3} gives one more example that also uses matte modulation.

\begin{figure*}[ht]
\centering
\begin{tabular}{cccc}
\includegraphics[width=0.24\linewidth]{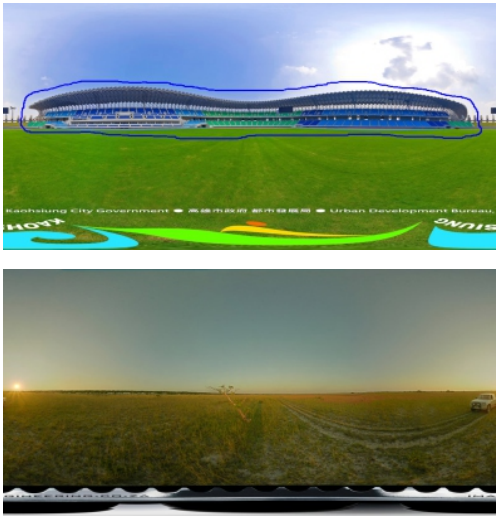}&\hspace{-0.25cm}\includegraphics[width=0.24\linewidth]{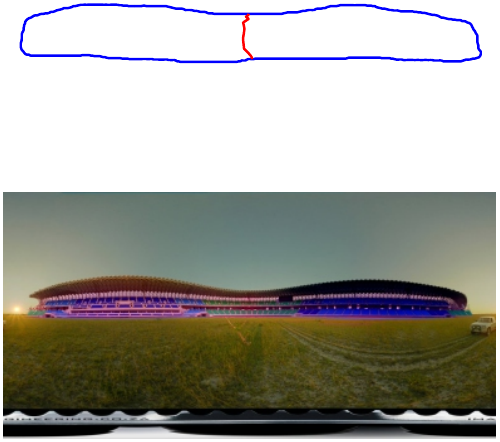}
&\hspace{-0.25cm}\includegraphics[width=0.24\linewidth]{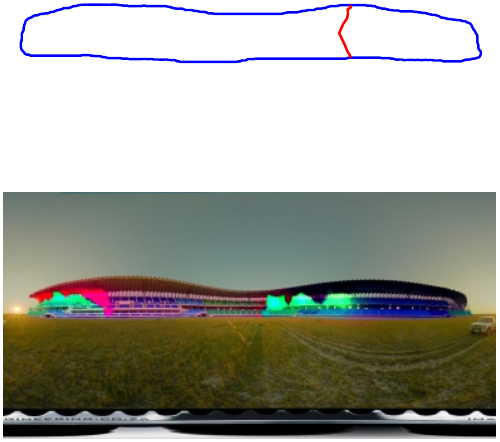}&\hspace{-0.25cm}\includegraphics[width=0.24\linewidth]{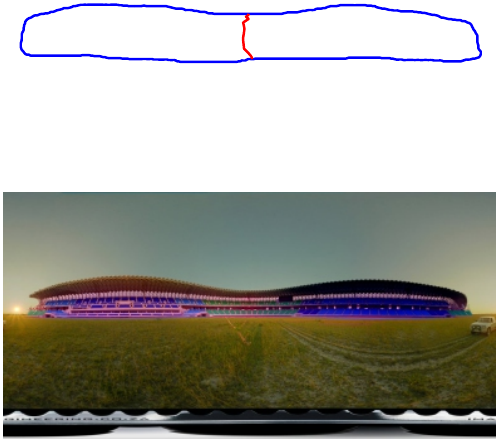}\\
(a)&(b)&(c)&(d)\\
\end{tabular}
\caption{The splitting paths and corresponding cloning results: (a) the source and target panoramas, (b) the splitting path from the method given in Section \ref{sec:path}, (c) the splitting path computed by performing PCA on the boundary of the spherical polygon (Alternative 1) and (d) the splitting path computed by performing PCA on the boundary of the projected spherical polygon (Alternative 2).}
\label{fig:pathExample}
\end{figure*}

\if 0
For the three examples shown in Figure \ref{fig:result3}, the source and target panoramas both have distinguishable skylines and the large patches covering more than 180$^\circ$ filed-of-view are all cloned at their default positions, i.e. the positions where they are at in the source images. In this case, the cloned objects are consistent with the surrounding of the target images and give pleasant result. However when the skylines in the target images are indistinguishable, simply cloning the patches at their default positions will give unreasonable results, as shown in Figure \ref{fig:result4}(a). For our panorama cloning, the problem can be solved by specifying carefully selected target cloning positions. The cloned objects will change their shapes according to the deformation of the target image as shown in Figure \ref{fig:result4}(b). However this approach will not work for the planar image cloning method (Figure \ref{fig:result4}(c)). This is because when the selected positions are different with the default positions, the source patch and the target patch will have different spherical deformations.
\fi

\subsection{Comparing Different Splitting Path Solutions}\label{sec:split}

In the experiments, we observe that the computed coordinates are not equal for different splitting paths. Besides the splitting path generation method given in Section \ref{sec:path}, we consider another two possible solutions and make comparison in the following.

\begin{itemize}
\item Alternative 1 performs the PCA on the vertices of the \textit{spherical polygon} $\mathrm{P}$. The first principal component is treated as the direction of the spherical polygon, and induces a great circle, whose normal vector is the principal component. Given the great circle, the splitting path then can be computed as the method given in Section \ref{sec:path}.
\item Alternative 2 performs the PCA on the \textit{projected spherical polygon} $\bar{\mathrm{P}}$, and the principal component is used to determine a great circle, based on which the splitting path is computed.
\end{itemize}

The generated splitting paths and their corresponding cloning results are show in Figure \ref{fig:pathExample}. We find when the range of the angles that the spherical polygon covers is considerably large, Alternative 1 may give inappropriate splitting path. As demonstrated in Figure \ref{fig:pathExample}(c), one sub-region after splitting still covers more than 180$^\circ$, and so the cloned result suffers from discoloration artifacts. On the other hand, Alternative 2 works well even for the cloned patches that are not horizontal. In our experience, the slanted objects that cover more than 180$^\circ$ are rare. Therefore we choose the method given in Section \ref{sec:path}, which avoid the PCA operation and is more timing efficient.

%\if 0
%As explained in Section \ref{sec:path}, different splitting path computation methods may give different cloning results. To further explore the impacts of the splitting path on the cloning result, we manually specify the start and end node of the path, and use these nodes to determine the splitting path. Fig.~\ref{fig:pathExample} (e)-(h) give four examples of different splitting pathes and the corresponding cloning results. Among the four pathes, the first two are on the left of the path given in Fig.~\ref{fig:pathExample} (b), the last two are on the right. The coverage of the two sub-region splitted by the path given in Fig.~\ref{fig:pathExample} (f) and (g) are all smaller than 180$^\circ$, and the cloning results do not suffer from the discoloration artifacts. However the results given in Fig.~\ref{fig:pathExample} (e) and (h) are not immune to this problem. This is because one of the sub-region still cover more than 180$^\circ$. So there is an interval, falling in which the splitting path will give the results that have no discoloration. And the range of the interval is inversely proportional to the coverage of the original spherical polygon. That is to say we can not find a reasonable splitting path for some cloned patches, e.g. the region covers nearly 360$^\circ$.
%\fi

\subsection{Timing Performance}

We test the timing performance on a desktop installed with Intel(R) Core(TM) i5¨C3330 CPU @ 3.00GHz and NVIDIA GeForce GT 640.
Our current implementation constructs the adaptive mesh, computes the spherical mean value coordinate, and evaluates the membrane at each mesh vertex on the CPU. The interpolation of pixels within each triangle of the mesh is performed on the GPU by exploiting the hardware rasterization ability.

\begin{table}[ht] \small
\renewcommand{\arraystretch}{1.3}
\caption{Performance statistics for panorama cloning: the times for splitting-based cloning are in bold type.}
\label{tbl:performance}
\centering
\begin{tabular}{|r|r|r|r|r|}
  \hline
  boundary & mesh & cloned & preprocessing & cloning \\
  pixels & vertices & pixels & time(ms) & time(ms) \\
  \hline
  \hline
  279& 1031 &  15009 &   23 & 0.342 \\ %no split g_iWidth/4
  \hline
  500& 1957 &  43955 &   71 & 0.387 \\ %no split g_iWidth/4
  \hline
  \multirow{2}{*}{691}& \multirow{2}{*}{2763} &  \multirow{2}{*}{77631} &  140 & 0.428 \\ %no split g_iWidth/4
  \cline{4-5}
  & & & \textbf{5873} & \textbf{0.434} \\ %   split g_iWidth/4
  \hline
  \multirow{2}{*}{900}&   \multirow{2}{*}{3661} &    \multirow{2}{*}{96989} &  238 & 0.472 \\ %no split g_iWidth/4
  \cline{4-5}
  & & & \textbf{9804} & \textbf{0.469} \\ %   split g_iWidth/4
  \hline
  \multirow{2}{*}{1179}& \multirow{2}{*}{4925} & \multirow{2}{*}{186314} &  419 & 0.531 \\ %no split g_iWidth/4
  \cline{4-5}
  & & &\textbf{17852} & \textbf{0.535} \\ %   split g_iWidth/4
  \hline
\end{tabular}
\end{table}

The timing is recorded as two parts: the preprocessing time, which correspond to the steps of adaptive mesh construction, spherical mean value coordinates computation and membrane evaluation; and the cloning time, which corresponds to pixel color interpolation. For a specified cloning patch, the preprocessing procedure is carried out only once for different target cloning positions. Table \ref{tbl:performance} gives the running time with respect to different number of boundary pixels, adaptive mesh vertices and cloned pixels. For the 3rd to 5th cases, which clone the patches that cover more than 180$^\circ$, we record the time for cloning without and with splitting (bold type). From the table we can see that as the number of boundary pixels and adaptive mesh vertices increases, more time is required for the preprocessing procedure. Although the cloning time increases as the cloned regions become larger, this part of time is relatively short. Another fact is that cloning with splitting takes much more preprocessing time than that without splitting, which seriously affects the timing performance.

Note that because panoramas are oversampled near poles, if we clone a patch near the equator of the source panorama to the polar regions of the target panorama, the cloned patch can be obviously blurred. Supersampling is used to deal with this problem, and Table \ref{tbl:performance2} shows the cloning times for different supersampling configurations when there are about 510 boundary pixels and 1970 adaptive mesh vertices. The cloning time without supersampling is the same to that of the 2nd case in Table \ref{tbl:performance}, in which the test cases are performed without supersampling. Table \ref{tbl:performance2} shows that the cloning time for 2$\times$2 supersampling is approximate to that without supersampling. When more samples are used, the cloning time is increased accordingly. Although it needs about 5ms to render a frame for 16$\times$16 supersampling, it can still clone the patch in real time. Actually for quality, the results shown in this paper are generated using this configuration.

\begin{table}[!h]\small
\renewcommand{\arraystretch}{1.3}
\caption{Performance statistics for different supersampling configuration}
\label{tbl:performance2}
\centering
\begin{tabular}{|l||r|r|r|r|r|}
  \hline
  supersampling & 1$\times$1 & 2$\times$2 & 4$\times$4 & 8$\times$8 & 16$\times$16\\
  \hline
  cloning time(ms)& 0.387 & 0.391 & 0.650 & 1.687 & 5.055\\
  \hline
\end{tabular}
\end{table}

\section{Conclusion}\label{sec:Conclusion}
In this paper, we propose a coordinate based algorithm to solve the problem of 360$^\circ$ panorama cloning, which has not been studied previously. In our work, we re-derive spherical mean value coordinates (SMVCs), and use SMVCs as the weights to achieve panorama cloning. To preserve the orientation of the cloned patch, we develop a two-step rotation computation method. A splitting-based method is proposed to remove discoloration artifacts for the case when cloning large patches cover more than 180$^\circ$ filed-of-view. With the proposed method, we can get satisfactory results and achieve real time cloning performance. As one future improvement, we would like to port the computation of SMVC and the evaluation of mean value coordinate membrane to the GPU.

% texture, color difference
%Our splitting-based cloning method is devised to remove the discoloration artifacts due to the region coverage. Similar artifacts may appear when there are large texture or color differences between the source and target patches. To handle such artifacts, the hybrid blending approach \cite{Chen2009}\cite{Chen2013} can be used, which classifies the polygon boundary into consistent part and inconsistent part based on color and texture difference.

%-------------------------------------------------------------------------
\appendix \label{Appendix:SMVC}
\numberwithin{equation}{section}
\section{Spherical Mean Value Coordinates}\label{sec:deduction}

The basic idea to compute spherical mean value coordinates of point $\mathbf{v}$ is first projecting the spherical polygon $\mathrm{P}$ to the tangent plane of unit sphere at $\mathbf{v}$ to get a planar polygon $\bar{\mathrm{P}}$, then computing planar mean value coordinates of $\mathbf{v}$ respect to $\bar{\mathrm{P}}$, followed by an anisotropic scaling. To project the spherical polygon, Langer et al.~\cite{SMVC_2006} use gnomonic projection, which would project two antipodal points to the same point on the tangent plane. In our work, we use the \emph{generalized stereographic projection} to deduce the spherical mean value coordinates. The main advantage of stereographic projection over gnomonic projection is that except for the projection point, it can project the entire sphere onto a plane without introducing overlapping. Particularly, this new deduction is helpful to analysis the decoloration problem.

%Because of this advantage, we also use stereographic projection to construct the adaptive mesh (Section \ref{sec:mesh}).
As illustrated in Figure~\ref{fig:SMVC}, we first move the origin of the 3D coordinate system from the sphere center to the projection point, which is the antipodal point of $\mathbf{v}$ and noted as
\begin{equation}
\dot{\mathbf{v}}=-\mathbf{v}.
\end{equation}
In the translated coordinate system, the boundary vertices $\mathbf{v}_{i}$ of spherical polygon, its projected counterpart $\bar{\mathbf{v}}_{i}$ and point $\mathbf{v}$ are represented by
\begin{equation}\label{equ:trans}
\mathbf{v}'_{i}=\mathbf{v}_{i}-\dot{\mathbf{v}}, \bar{\mathbf{v}}'_{i}=\bar{\mathbf{v}}_{i}-\dot{\mathbf{v}}, \mathbf{v}'=\mathbf{v}-\dot{\mathbf{v}}.
\end{equation}

Following 2D mean value coordinate definition, the planar mean value coordinates of $\mathbf{v}'$ with respect to the projected polygon $\bar{\mathrm{P}}'=\{\bar{\mathbf{v}}'_{1},\ldots,\bar{\mathbf{v}}'_{n}\}$ are given by
\begin{equation}\label{equ:barLambda}
\bar{\lambda}_{i} = \frac{w_{i}}{\sum_{j}w_{j}}, \; w_{i} = \frac{\tan\frac{\alpha_{i-1}}{2}+\tan\frac{\alpha_{i}}{2}}{d_{i}},
\end{equation}
where $\alpha_{i}$ is the signed angle between vectors $\mathbf{v}'\times\mathbf{v}'_{i}$ and $\mathbf{v}'\times\mathbf{v}'_{i+1}$, and the distance $d_{i}=\|\bar{\mathbf{v}}'_{i}-\mathbf{v}'\|$ as shown in Figure~\ref{fig:SMVC}. Because each $\bar{\mathbf{v}}'_{i}$ is a scaling of $\mathbf{v}'_{i}$, the coordinate $\tilde{\lambda}_{i}$ of $\mathbf{v}'$ with respect to the spherical polygon $\mathrm{P}'=\{\mathbf{v}'_{1},\ldots,\mathbf{v}'_{n}\}$ (in the translated coordinate system) can be constructed from $\bar{\lambda}_{i}$, i.e.,
\begin{equation}\label{equ:tildeLambda}
\tilde{\lambda}_{i}=\frac{\|\bar{\mathbf{v}}'_{i}\|}{\|\mathbf{v}'_{i}\|}\bar{\lambda}_{i}=\frac{2}{1+\cos\theta_{i}}\bar{\lambda}_{i},
\end{equation}
where $\theta_{i}$ is the angle between $\mathbf{v}$ and $\mathbf{v}_{i}$. What's more, the coordinate $\tilde{\lambda}_{i}$ defined in Equation~\ref{equ:tildeLambda} satisfies
\begin{equation}\label{equ:tildeLambdaLinear}
\sum\nolimits_{i}\tilde{\lambda}_{i}\mathbf{v}'_{i}=\mathbf{v}'.
\end{equation}

Note that the current computation is operated in the translated coordinate system (originated at $-\mathbf{v}$), while our goal is computing the spherical mean value coordinate $\lambda_{i}$ of point $\mathbf{v}$ with respect to the spherical polygon $\mathrm{P}$ (in the original coordinate system centered at sphere center), which should satisfy
\begin{equation}\label{equ:lambdaLinear}
\sum\nolimits_{i}\lambda_{i}\mathbf{v}_{i}=\mathbf{v}.
\end{equation}
Substituting $\mathbf{v}'_{i}$ and $\mathbf{v}'$ with Equation~\ref{equ:trans}, Equation~\ref{equ:tildeLambdaLinear} becomes
\begin{equation}
\sum\nolimits_{i}\tilde{\lambda}_{i}(\mathbf{v}_{i}+\mathbf{v})=2\mathbf{v},
\end{equation}
which can be further rewritten as
\begin{equation}
\sum\nolimits_{i}\tilde{\lambda}_{i}\mathbf{v}_{i}=(2-\sum\nolimits_{i}\tilde{\lambda}_{i})\mathbf{v}.
\end{equation}
By comparing Equation~\ref{equ:lambdaLinear} with the above equation, we can get
\begin{equation}\label{equ:lambda}
\lambda_{i} = \frac{\tilde{\lambda}_{i}}{2-\sum_{i}\tilde{\lambda}_{i}}.
\end{equation}
By using Equation~\ref{equ:barLambda},~\ref{equ:tildeLambda} and \ref{equ:lambda} in addition to some trigonometrics identities, the spherical mean value coordinates will have the final form as
\begin{equation}
\lambda_{i}(\mathbf{v})=\frac{(\tan\frac{\alpha_{i-1}}{2}+\tan\frac{\alpha_{i}}{2})/\sin\theta_{i}}{\sum\nolimits_{j}\cot\theta_{j}(\tan\frac{\alpha_{j-1}}{2}+\tan\frac{\alpha_{j}}{2})}.
\end{equation}

%It is worthy mentioning that our deduction is more straightforward as compared to the work in \cite{SMVC_2006}. Another more important advantage of our deduction is that the analysis of decoloration artifacts in Section~\ref{sec:overfit} will benefit from the use of Eq. (\ref{equ:tildeLambda}) and Eq. (\ref{equ:lambda}).

%-------------------------------------------------------------------------
\bibliographystyle{eg-alpha}

\bibliography{SMVC}

\newcommand{\etalchar}[1]{$^{#1}$}
\begin{thebibliography}{\uppercase{PBDSH13}}

\bibitem[ADF{\etalchar{*}}10]{StreetView}
\textsc{Anguelov D., Dulong C., Filip D., Frueh C., Lafon S., Lyon R., Ogale
  A., Vincent L., Weaver J.}:
\newblock {Google Street View}: Capturing the world at street level.
\newblock \emph{Computer 43}, 6 (2010), 32--38.

\bibitem[AH15]{MPB2015}
\textsc{Afifi M., Hussain K.~F.}:
\newblock {MPB}: A modified poisson blending technique.
\newblock \emph{Computational Visual Media 1}, 4 (2015), 331--341.

\bibitem[B\"04]{Harmonic04}
\textsc{B\"{u}low T.}:
\newblock Spherical diffusion for 3d surface smoothing.
\newblock \emph{IEEE Transactions on Pattern Analysis and Machine Intelligence
  26}, 12 (2004), 1650--1654.

\bibitem[BWS{\etalchar{*}}13]{Intent13}
\textsc{Bie X., Wang W., Sun H., Huang H., Zhang M.}:
\newblock Intent-aware image cloning.
\newblock \emph{The Visual Computer 29}, 6 (2013), 599--608.

\bibitem[CCT{\etalchar{*}}09]{Chen2009}
\textsc{Chen T., Cheng M.-M., Tan P., Shamir A., Hu S.-M.}:
\newblock Sketch2photo: Internet image montage.
\newblock In \emph{ACM SIGGRAPH Asia} (2009), pp.~124:1--124:10.

\bibitem[CZSH13]{Chen2013}
\textsc{Chen T., Zhu J.-Y., Shamir A., Hu S.-M.}:
\newblock Motion-aware gradient domain video composition.
\newblock \emph{IEEE Transactions on Image Processing 22}, 7 (2013),
  2532--2544.

\bibitem[FHL{\etalchar{*}}09]{MVC_Cloning_2009}
\textsc{Farbman Z., Hoffer G., Lipman Y., Cohen-Or D., Lischinski D.}:
\newblock Coordinates for instant image cloning.
\newblock In \emph{ACM SIGGRAPH} (2009), pp.~67:1--67:9.

\bibitem[Flo03]{Floater03}
\textsc{Floater M.~S.}:
\newblock Mean value coordinates.
\newblock \emph{Computer Aided Geometric Design 20}, 1 (2003), 19 -- 27.

\bibitem[HF06]{MVC_ToG_2006}
\textsc{Hormann K., Floater M.~S.}:
\newblock Mean value coordinates for arbitrary planar polygons.
\newblock \emph{ACM Transactions on Graphics 25}, 4 (2006), 1424--1441.

\bibitem[JSTS06]{DragDrop06}
\textsc{Jia J., Sun J., Tang C.-K., Shum H.-Y.}:
\newblock Drag-and-drop pasting.
\newblock In \emph{ACM SIGGRAPH} (2006), pp.~631--637.

\bibitem[JSW05]{MVCfCTM2005}
\textsc{Ju T., Schaefer S., Warren J.}:
\newblock Mean value coordinates for closed triangular meshes.
\newblock \emph{ACM Trans. Graph. 24}, 3 (2005), 561--566.

\bibitem[KH10]{Metric_2010}
\textsc{Kazhdan M., Hoppe H.}:
\newblock Metric-aware processing of spherical imagery.
\newblock In \emph{ACM SIGGRAPH Asia} (2010), pp.~149:1--149:10.

\bibitem[LBS06]{SMVC_2006}
\textsc{Langer T., Belyaev A., Seidel H.-P.}:
\newblock Spherical barycentric coordinates.
\newblock In \emph{Proceedings of the Fourth Eurographics Symposium on Geometry
  Processing} (2006), pp.~81--88.

\bibitem[LSC{\etalchar{*}}12]{SteroCloning}
\textsc{Luo S.-J., Shen I.-C., Chen B.-Y., Cheng W.-H., Chuang Y.-Y.}:
\newblock Perspective-aware warping for seamless stereoscopic image cloning.
\newblock In \emph{ACM SIGGRAPH Asia} (2012), pp.~182:1--182:8.

\bibitem[MK09]{Reconstruction}
\textsc{Micusik B., Kosecka J.}:
\newblock Piecewise planar city 3d modeling from street view panoramic
  sequences.
\newblock In \emph{CVPR} (2009), pp.~2906--2912.

\bibitem[MP08]{GPUPoisson}
\textsc{McCann J., Pollard N.~S.}:
\newblock Real-time gradient-domain painting.
\newblock In \emph{ACM SIGGRAPH} (2008), pp.~93:1--93:7.

\bibitem[PBDSH13]{WAoS2013}
\textsc{Panozzo D., Baran I., Diamanti O., Sorkine-Hornung O.}:
\newblock Weighted averages on surfaces.
\newblock \emph{ACM Trans. Graph. 32}, 4 (2013), 60:1--60:12.

\bibitem[PGB03]{PoissonEditing03}
\textsc{P{\'e}rez P., Gangnet M., Blake A.}:
\newblock Poisson image editing.
\newblock In \emph{ACM SIGGRAPH} (2003), pp.~313--318.

\bibitem[RC14]{Floorplan14}
\textsc{Ricardo~Carbral Y.~F.}:
\newblock Piecewise planar and compact floorplan reconstruction from images.
\newblock In \emph{CVPR} (2014).

\bibitem[Rus10]{BCoS2010}
\textsc{Rustamov R.~M.}:
\newblock Barycentric coordinates on surfaces.
\newblock \emph{Computer Graphics Forum 29}, 5 (2010), 1507--1516.

\bibitem[SS95]{Wavelets95}
\textsc{Schr\"{o}der P., Sweldens W.}:
\newblock Spherical wavelets: Texture processing.
\newblock Eurographics. 1995, pp.~252--263.

\bibitem[TBP13]{ChangeDetection13}
\textsc{Taneja A., Ballan L., Pollefeys M.}:
\newblock City-scale change detection in cadastral 3d models using images.
\newblock In \emph{CVPR} (2013), pp.~113--120.

\bibitem[WCPB10]{Harmonic10}
\textsc{Wang R., Chen W.-f., Pan M.-h., Bao H.-j.}:
\newblock Harmonic coordinates for real-time image cloning.
\newblock \emph{Journal of Zhejiang University SCIENCE C 11}, 9 (2010),
  690--698.

\bibitem[XEOT12]{SUN360}
\textsc{Xiao J., Ehinger K., Oliva A., Torralba A.}:
\newblock Recognizing scene viewpoint using panoramic place representation.
\newblock In \emph{CVPR} (2012), pp.~2695--2702.

\bibitem[XSMC10]{Xie2010}
\textsc{Xie Z.-F., Shen Y., Ma L.-Z., Chen Z.-H.}:
\newblock Seamless video composition using optimized mean-value cloning.
\newblock \emph{The Visual Computer 26}, 6 (2010), 1123--1134.

\bibitem[ZMH15]{PCfSV2015}
\textsc{Zhu Z., Martin R.~R., Hu S.-M.}:
\newblock Panorama completion for street views.
\newblock \emph{Computational Visual Media 1}, 1 (2015), 49--57.

\bibitem[ZT11]{Zhang2011}
\textsc{Zhang Y., Tong R.}:
\newblock Environment-sensitive cloning in images.
\newblock \emph{The Visual Computer 27}, 6 (2011), 739--748.

\bibitem[ZWF{\etalchar{*}}13]{Cube2Video13}
\textsc{Zhao Q., Wan L., Feng W., Zhang J., Wong T.-T.}:
\newblock Cube2video: Navigate between cubic panoramas in real-time.
\newblock \emph{IEEE Transactions on Multimedia 15}, 8 (2013), 1745--1754.

\end{thebibliography}

\end{document}